\documentclass[aip,
 amsmath,amssymb,
 nofootinbib,
reprint,%
 Conference Proceedings,
]{revtex4-1}

\usepackage{graphicx}
\usepackage{dcolumn}
\usepackage{bm}
\usepackage[utf8]{inputenc}
\usepackage[T1]{fontenc}
\usepackage{mathptmx}
\usepackage{etoolbox}
\usepackage{mathtools}
\usepackage{color}
\usepackage{xcolor}
\usepackage{calc}
\usepackage{relsize}
\usepackage{multirow}
\usepackage{rotating}
\usepackage{url}
\usepackage{braket}
\usepackage{slashed}
\usepackage[newcommands]{ragged2e}
\usepackage{amsfonts}
\usepackage{empheq}
\usepackage{tikz}
\usepackage{subcaption}
\usetikzlibrary{decorations.pathmorphing,decorations.markings}
\tikzset{snake it/.style={decorate, decoration=snake}}
\usepackage{upgreek}
\usepackage{scalerel}
\usepackage[font=small,labelfont=bf,
   justification=Justified,
   format=plain]{caption}

\newcommand{\be}{\begin{equation}}
\newcommand{\ee}{\end{equation}}
\newcommand{\bi}{\begin{enumerate}}
\newcommand{\ei}{\end{enumerate}}
\newcommand{\ud}{{\mathrm{d}}}
\hfuzz=\maxdimen \tolerance=10000
\vfuzz=\maxdimen \tolerance=10000
\newcommand{\LCm}{{\scriptscriptstyle -}}
\newcommand{\LCp}{{\scriptscriptstyle +}}
\newcommand{\LCpm}{{\scriptscriptstyle \pm}}

\newcommand{\LCperp}{{\scriptscriptstyle \perp}}
\renewcommand{\j}{\theta}
\newcommand{\e}{\mathrm{e}}
\newcommand{\nroots}{\jbar{N}}
\newcommand*\jbar[1]{%
  \hbox{%
    \vbox{%
      \hrule height 0.1pt 
      \kern0.5ex
      \hbox{%
        \kern-0.15em
        \ensuremath{#1}%
        \kern-0.02em
      }%
    }%
  }%
} 
\newcommand{\llangle}{\langle\!\langle}
\newcommand{\rrangle}{\rangle\!\rangle}
\def\ddel{{}^\bullet\! \Delta}
\def\deld{\Delta^{\hskip -.5mm \bullet}}

\def\ddeld{{}^{\bullet}\! \Delta^{\hskip -.5mm \bullet}}

\newcommand{\Pn}{\bar{\mathfrak{P}}}
\newcommand{\Del}{\underset{\smile}{\Updelta}}
\newcommand{\dDel}{{}^{\bullet}\!\Del{}}

\newcommand{\oDel}{{}^{\circ}\!\Del{}}

\newcommand{\oDelo}{{}^{\circ}\!\Del\!{}^{\circ}\!}

\newcommand{\G}{\mathcal{G}}

\DeclareMathOperator{\tr}{tr}

\newcommand{\tri}{\scalebox{0.6}{$\triangle$}}

\makeatletter
\def\@email#1#2{%
 \endgroup
 \patchcmd{\titleblock@produce}
  {\frontmatter@RRAPformat}
  {\frontmatter@RRAPformat{\produce@RRAP{*#1\href{mailto:#2}{#2}}}\frontmatter@RRAPformat}
  {}{}
}%
\makeatother
\begin{document}

\preprint{AIP/123-QED}

\title[Worldline Modeling of Ultra-Intense Lasers for N-photon Scattering Processes]{Worldline Modeling of Ultra-Intense Lasers for N-photon Scattering \\Processes}
\author{I. Ahumada}
\author{P. Copinger}%
 \email{patrick.copinger@plymouth.ac.uk}
\author{J. P. Edwards}
\affiliation{ 
Centre for Mathematical Sciences, University of Plymouth, Plymouth, PL4 8AA, UK}
\author{K. Rajeev}
\affiliation{Higgs Centre, School of Physics and Astronomy, University of Edinburgh, EH9 3FD, UK}

\date{\today}

\begin{abstract}
The modeling of present and future ultra-intense lasers demands techniques that go beyond the standard diagrammatic approach to non-perturbatively fully capture the effects of strong fields. We illustrate the first-quantized path integral representation for strong-field quantum electrodynamics as a means of accessing the laser being treated as a background field, which is treated without  recourse to perturbation theory. We examine an all-multiplicity construction for $N-$photon scattering processes for complex scalars and spinors, showing compact Master Formulae for tree-level scattering. Several background fields are considering including: plane waves, impulsive PP-waves, non-null fields, and homogeneous fields (constant-crossed fields) with low-energy external photons.
\end{abstract}

\maketitle

\section{\label{sec:intro}Introduction}

Modern facilities equipped to produce ultra-intense lasers in the multi-petawatt regime are on the cusp of directly probing the physics of strong field quantum electrodynamics (SFQED)~\cite{PhysRevLett.76.3116,PhysRevX.8.011020,PhysRevX.8.031004}. The strong electromagnetic fields that are present in ultra-intense lasers~\cite{Strickland:1985gxr} have particle interactions that induce nonlinear and nonperturbative effects, which are challenging to capture theoretically using a conventional diagrammatic approach. The field strength at which these effects become important, signalling the regime of SFQED, is characterized by the Schwinger critical field, which occurs at $m^2/e=1.3\times 10^{16}\, \text{V/m}$ for $m$ the particle mass~\cite{Fedotov:2022ely}. And while the laser intensity associated to the Schwinger critical field still has not been directly reached in a laser~\cite{Yoon:21}, the intensity seen by oncoming particles in an electron beam colliding with a laser in a Lorentz boosted frame can surpass the critical intensity. Thus, modern laser facilities are actively probing with optical light the nonlinear effects of SFQED~\cite{LUXE:2023crk,PhysRevAccelBeams.22.101301, Ahmadiniaz:2024xob,CMS,ATLAS}, and there is demand from theory to supply relevant measurable quantities.

Despite our understanding of quantum electrodynamics (QED), evidenced through high-accuracy measurements of e.g. the anomalous magnetic moment~\cite{Laporta4}, our understanding of SFQED especially at higher loop and multiplicity order is still in development, impacting experimental verification. This is because the problems are difficult; for example in a plane wave background the state-of-the-art is at 4-point scattering at \textit{tree} level~\cite{Fedotov:2022ely}. Further the perturbative breakdown of SFQED at high intensities, in what is known as the Ritus-Narozhnyi conjecture~\cite{RitusRN1,Narozhnyi:1980dc,Fedotov:2016afw}, is thought to require an all orders resummation in order to make meaningful predicts. New techniques to address these outstanding theoretical shortcomings are needed. Here we report on recent advances towards the study of high multiplicity in ultra-intense laser targeted backgrounds using the worldline formalism. Indeed, nonlinear Compton scattering of hundreds of optical photons with GeV electrons was observed against locally constant field approximation (LCFA) predictions~\cite{2024NaPho..18.1212M}, making the high multiplicity regime imminently accessible. Let us also mention that the worldline formalism is but one approach for multi-photon scattering processes; for example one may use a diagrammatic approach in the Furry picture using background field dressed propagators, etc. interacting with any number of $N-$photons.

The worldline formalism, first reported by Feynman~\cite{Feyn1,Feyn2}, but had regained popularity following the work of several others~\cite{Bern:1990cu,Bern:1991aq,Strass1}, is a first quantized approach to quantum field theory, in particular SFQED. Its usage has proved instrumental for the Schwinger effect~\cite{PhysRev.82.664}, vacuum birefringence~\cite{AntonPlane}, nonlinear Breit-Wheeler pair production~\cite{DegliEsposti:2021its}, and Berry's phase in quantum field theory (QFT)~\cite{Copinger:2022jgg}. And it is by construction well-suited to nonperturbative studies, by e.g. worldline instantons~\cite{Affleck:1981bma,Dunne:2005sx,Dunne:2006st,Dumlu:2011cc,Ilderton:2015qda,Lan:2018xnq,Copinger:2020feb}. Non-linear effects have also been analyzed in the propertime setting~\cite{doi:10.1142/S0217751X04017744}.
Furthermore, and the focus of this paper, is the extension of the worldline formalism to incorporate external photon legs onto the loop and line. At arbitrary $N-$photon multiplicity these are referred to as `Master Formulae.' This was first established in the vacuum~\cite{Strass1,Schmidt:1993rk,103}, and has since been applied to the cases of homogenous electromagnetic fields~\cite{McKeon:1994hd,Shaisultanov:1995tm,Adler:1996cja,Reuter:1996zm}, and plane waves~\cite{Edwards:2021vhg, Edwards:2021uif,Schubert:2023gsl}. More recently the worldline Master Formulae were connected through Lehmann–Symanzik–Zimmermann (LSZ) onto the matter line to study amplitudes~\cite{Bonocore:2020xuj, Mogull:2020sak}. In this work we report on recent progress on Master Formulae and their corresponding amplitudes as are relevant for the modeling of ultra-high lasers as a background field.

\section{\label{sec:back}Background field models}

We outline several target background fields here, that will then be used in the $N-$photon worldline formalism to furnish us with an all multiplicity Master Formulae, from which using LSZ scattering amplitudes can be derived. Several background fields are discussed including plane waves, impulsive PP-waves, non-null backgrounds, and homogeneous backgrounds coupled to low-energy photons.

\paragraph{Plane waves:}
The first background we consider is naturally the case of a plane wave. The Klein-Gordon and Dirac equations admit exact solutions in a place wave background~\cite{Fradkin:1991zq}; thus is it unsurprising that exact solutions in the worldline formalism have been well-studied~\cite{Edwards:2021vhg, Edwards:2021uif,Schubert:2023gsl}. To maintain relevance to ultra-intense lasers we allow our background field to take on any amplitude and wave vector. For concreteness we consider a plane wave with gauge
\be\label{planewave}
e A_\mu(x) = a_\mu(x^\LCp) = \delta_\mu^\LCperp a_\LCperp (x^\LCp)\,,\quad a_\LCperp(\infty)= a_\LCperp^\infty\,.
\ee
We work in light front coordinates with $\ud s^2 = \ud x^\LCp \ud x^\LCm - \ud x^\LCperp \ud x^\LCperp$, where $x^\LCperp = (x^1,x^2)$ and our light front time is characterized with null vector $n^\mu$ such that $x^\LCp=x\cdot n$. Other backgrounds to follow also adhere to these light front coordinate conventions.

\paragraph{Impulsive PP-waves:}
Next, we consider the case of an impulsive PP-wave, which we study under the following gauge vanishing at asymptotic light front time
\be\label{shockwave}
eA_{\mu}(x)=-en_{\mu}\delta(x^\LCp)\Phi(x^{\LCperp})\,,
\ee
where $\Phi(x^{\LCperp})$ is taken to be an arbitrary function~\cite{Adamo:2021jxz}. They are obtained from ultra-boosted arbitrary source distributions. Impulsive PP-waves are a natural generalization of shockwaves; one need only take $\Phi(x^\LCperp)\propto \log(\mu^2|x^{\LCperp}|^2)$ with $\mu$ constant to find a shockwave solution. Impulsive plane waves too are incorporated in the above; here one can write $\Phi(x^\LCperp) = r_\LCperp x^\LCperp$ with $r_\LCperp$ a constant.

\paragraph{Non-null background:}
Next, we consider as an immediate generalization to the case of a plane wave,~\eqref{planewave}, a similar gauge structure, however with dependence not in light front time but rather on a light front coordinate that is not null. We label this coordinate as $x^{\triangle} = x^{+} + \frac{\rho^{2}}{4} x^\LCm=\tilde{n}\cdot x$ with $\tilde{n}^2=\rho^2\neq 0$, which represents a rotation away from $x^\LCp$ to $x^\LCp$ and $x^\LCm$. And our background gauge field is
\be\label{nonnull}
e A_\mu(x) = a_\mu(x^{\tri}) = \delta_\mu^\LCperp a_\LCperp (x^{\tri})\,,\quad a_\LCperp(\infty)= a_\LCperp^\infty\,.
\ee
Such a configuration is useful for more practical background field modeling to go beyond the plane wave background. Apart from the plane wave case, $\rho^2=0$, two special cases are possible: 1) $\tilde{n}^2<0$ corresponds to the case of an undulator background~\cite{Heinzl:2016kzb}, or rather one of a plane wave in a refractive index greater than one~\cite{CRONSTROM1977137}. 2) $\tilde{n}^2>0$ corresponds to a time-dependent electric field~\cite{Bulanov:2003aj}, or rather one of a plane wave in refractive index less than one~\cite{CRONSTROM1977137,BECKER1977601}.

\paragraph{Homogeneous background with low-energy photons:}
As a final example we examine a special combination of background field and external photons that dramatically simplifies the Master Formula. The background field is one of homogeneous fields, but with arbitrary strength. Here we write it in a Fock-Schwinger gauge as
\be\label{fs_homo}
    eA_\mu(x)=-\frac{1}{2}f_{\mu\nu}(x^\nu-\hat{x}^\nu)\,,
\ee
where $f_{\mu\nu}$ is a constant, and can be taken to be the field strength of a constant crossed field (CCF), $f_{\mu\nu}=n_\mu a_\nu-n_\nu a_\nu$; we however, leave it arbitrary. And $\hat{x}^\nu$ is a reference point. The trick we will go on to show is to couple to low energy photons~\cite{ChrisLow}. Namely, these are photons with energy $\omega_{i}  \ll m$ for all $i\in N$ that are on-shell with $k_{i}^{\mu} = (\omega_{i}, \omega_{i} \hat{\mathbf{k}}_{i})$. We further demand that $k_{i} \cdot k_{j} \ll m^2$ for $i \neq j$. What this ultimately entails is that one keep terms to linear order in all $k_i$. This we will show has the effect of turning the vertex operator into one of a homogeneous field term that is quadratic in the coordinates. We will show this simplification can be used to dramatically reduce the complexity of the Master Formula, as has been shown on the loop in vacuum and parallel fields~\cite{ChrisLow,Edwards:2018vjd,Ahmadiniaz:2023jwd,MishaLow}.

In the above four models, the physically relevant parameters observable in future experiments include~\cite{Fedotov:2022ely} the classical non-linearity parameter, $\xi=eE/mc\omega$, the energy of the collision between probe and the background, $\eta=\hbar k\cdot p/m^2 c^4$, and the quantum non-linearity parameter, $\chi=e\hbar\sqrt{-(F\cdot p)^2}/m^3c^4$ for probe momentum $p$, electron mass, $m$, background field strength $E$ or $F^{\mu\nu}$ and wave vector $k^\mu=\omega n^\mu$. In the below amplitudes and $N-$photon dressed propagators; the above parameters are all that may be seen. We have, however, elected to keep our expressions explicit in terms of probe (whether on the loop or line) or background field quantities for generality, and to maintain an exact form (the Schwinger propertime integral in most instances cannot be solved in closed form).

\section{\label{sec:worldline}$\bm{N-}$photon dressed worldline formalism}

\subsection{Complex scalars}
To begin our discussion of ultra-intense lasers for $N-$photon scattering processes we must first examine the worldline formalism. Consider, for example, a complex scalar theory without dynamical photons, but with the background field treated to all orders in the coupling. The propagator can be expressed in formal operator form as
$\mathcal{D}^{x'x}=-i \langle x' |(\hat{D}^2+m^2-i\epsilon)^{-1}|x\rangle$,
where the covariant derivative is $D^\mu=\partial^\mu +ia^\mu(x)$ for background field $a^\mu=eA^\mu$. Then, the Schwinger propertime representation~\cite{PhysRev.82.664} is arrived at by introducing a Laplace transform, exponentiating the inverse operator, through the introduction of an integral. Finally, one may express the resulting evolution operator in Schwinger propertime as a path integral over worldline trajectories as~\cite{Schubert:2001he,Edwards:2019eby}
\be
    \mathcal{D}^{x'x}[a] =
    \int_{0}^{\infty}\ud T \,\e^{-im^{2} T}\,
\int^{x(T)=x'}_{x(0)=x}\hspace{-1em}\mathcal{D}x(\tau)\, \e^{iS_{\mathrm{B}}[a]}\,,
\ee
with worldline action reading $S_{\mathrm{B}}[a]=-\int^T_0[\dot{x}/4+a(x)\cdot\dot{x}]$. 

Let us now consider the extension of the background field approach to incorporate $N-$external photons in an all-multiplicity construction. To do so one may repeat the same steps above for an $N-$photon dressed gauge plus background field as $a\to a+eA^\gamma$ with
\be\label{A_gamma}
    A^{\gamma}_{\mu}(x) = \sum_{i=1}^{N}\varepsilon_{\mu\, i}\e^{i k_{i}\cdot x}\,,
\ee
for polarization $\varepsilon_{\mu\, i}$ with corresponding photon momentum $k_i^\mu$. Then we take a multi-linear projection of the $N-$photon insertions to turn the $N-$photon dressed contribution into an expectation valued expression over worldline paths. We find, writing for the $N-$photon dressed case, for the complex scalar propagator
\be\label{D_N}
    \mathcal{D}^{x'x}_N[a] = 
    (-ie)^{N}
    \int_{0}^{\infty}\ud T \,\e^{-im^{2} T}\,
\int^{x(T)=x'}_{x(0)=x}\hspace{-2em}\mathcal{D}x(\tau)\, \e^{iS_{\mathrm{B}}[a]}\prod_{i=1}^{N} V^{x'x}[\varepsilon_{i}, k_i]\,.
\ee
The $N-$photons insertions now appear as vertices where we further simplify their subsequent evaluation through the use on a multi-linearization operator,
\begin{align}
    \prod_{i=1}^{N}V^{x'x}[\varepsilon_i,k_i] &=
    \prod_{i=1}^{N}\int_{0}^{T}d\tau_i\,\varepsilon_i\cdot\dot{x}(\tau_i)\, \e^{ik_i\cdot x(\tau_i)}\notag\\
    &=\prod_{i=1}^{N} \int_{0}^{T}\!\ud\tau_i\,
    \e^{-i\int^T_0d\tau\,\mathcal{J}\cdot x}
    \Big|_{\mathrm{lin}\,N}\,.
\end{align}
where $\mathcal{J}^{\mu}=i\sum_{i=1}^{N}\big(ik_{i}^\mu - \varepsilon_{i}^\mu\frac{\ud}{\ud\tau}\big)\delta(\tau-\tau_{i})$. The $\mathrm{lin}\,N$ dictates that we keep only the linear contribution of each $\varepsilon_i$ for $i=1,\cdots,N$ in the final expression. Equipped with the above form of the propagator, we may in a compact form study tree-level scattering on the matter line for \textit{any} number of photons.

\subsection{Spinors}
We will continue an analogous discussion for the case of spinors. Remarkably, apart from a subleading contribution, the only major difference between the scalar case is the inclusion of a spin factor, which we will show. First, however, let us write down the spinor propagator in its background field form, without yet an $N-$photon coupling. The propagator is
$\mathcal{S}^{x'x}[a]=(-i\slashed{D}_{x^\prime}+m)\mathcal{K}^{x'x}[a]$,
where the spacetime kernel reads
\be
    \mathcal{K}^{x'x}[a]=
    \int_{0}^{\infty} \!\ud T\, \e^{-im^2 T} \!
    \int_{x(0) = x}^{x(T) = x'}\hspace{-1.5em}\mathcal{D}x(\tau) \, \e^{i S_{\mathrm{B}}[a]}\,
    \mathrm{Spin}[a]\,,
\ee
where the spin factor is $\mathrm{Spin}[a]=\mathrm{symb}^{-1}\mathfrak{W}_\eta[a]$, and
\be
    \mathfrak{W}_\eta[a]={2^{-\frac{D}{2}}}\oint_{\textrm{A/P}}\hspace{-0.75em}\mathcal{D}\psi(\tau) \,   
    \e^{i\widetilde{S}_{\mathrm{F}}[a]}\,,
\ee
and whose spin factor action is $\widetilde{S}_{\mathrm{F}}[a] = \int_{0}^{T}\ud \tau[\frac{i}{2}\psi\cdot\dot{\psi} +i(\psi(\tau) + \eta) \cdot f(x(\tau)) \cdot (\psi(\tau) + \eta)]$. To account for the spin degrees of freedom, we use Grassmann variables in the spin factor who obey anti-periodic boundary conditions, i.e. $\psi(0)=-\psi(T)$. And to convert the spin factor into a Dirac matrix representation, we further utilize the propertime independent auxiliary variable, $\eta$, which is acted upon by the inverse symbolic map, $\mathrm{symb}^{-1}$. We refer the reader to Ref.~\onlinecite{Copinger:2023ctz} for details on how to apply the map, suffice to say, however, that for (3+1)-dimensions after the inverse map is applied only a maximum four gamma matrices would remain.

As before with the scalar propagator, we may write a propagator with $N-$photon insertions to study tree-level scattering. Applying the shift, as before, $a\to a+eA^\gamma$, with $A^\gamma$ represented by~\eqref{A_gamma}, we may decompose the spinor propagator with $N-$photons into both a sub-leading and a leading factor after taking the multi-linear projection
\be
    \mathcal{S}^{x'x}_N[a] =  
    (-i\slashed{\partial}_{x'}+\slashed{a}(x^{\prime \LCp})-m)    {\mathcal K}_{N}^{x'x}[a]+e\slashed{A}^\gamma(x') {\mathcal K}_{N-1}^{x'x}[a]\,.
\ee
The subleading factor is the second term proportional to $eA^\gamma$ and accounts for reduced external photon vertices $N-1$. What remains is the spacetime kernel whose form now closely resembles that of the complex scalar case but with a spin factor:
\begin{align}
    {\mathcal K}_{N}^{x'x}[a]&=
    (-ie)^{N}\int_{0}^{\infty} \!\ud T\, \e^{-im^2 T}\!
    \int_{x(0) = x}^{x(T) = x'}\hspace{-1.5em}\mathcal{D}x(\tau) \, \e^{i S_{\mathrm{B}}[a]}\,\notag\\
    &\times\mathrm{symb}^{-1}\mathfrak{W}[a]\prod_{i=1}^{N}V_{\eta}^{x'x}[\varepsilon_i,k_i]\,.
\end{align}
The spinor vertex now contains within it both the fluctuating Grassmann variable $\psi(\tau)$ and auxiliary Grassmann variable $\eta$ begin acted upon by the inverse symbolic map, which reads
\be
    \prod_{i=1}^{N}V_{\eta}^{x'x}[\varepsilon_i,k_i]
    =\prod_{i=1}^{N}
    \int_{0}^{T}\ud\tau_i\, \e^{-i\int^T_0\!d\tau\mathcal{J}\cdot x-i\sum_j^N
    \psi_\eta(\tau_{j}) \cdot \tilde{F}_{j}\cdot\psi_\eta(\tau_{j})}\Big|_{\mathrm{lin}\,N}\,.
\ee
Here the reduced and effective field strength of the external photons is $\tilde{F}^{\mu\nu}_i\coloneqq i[k^{\mu}_i\varepsilon^{\nu}_i-k^{\nu}_i\varepsilon^{\mu}_i]$, (we will also make use of $\tilde{f}_i=e\tilde{F}_i$), and we further write for brevity $\psi_\eta\coloneqq\psi+\eta$. It is further convenient to split up the vertex so that we may write
\begin{align}
    {\mathcal K}_{N}^{x'x}[a]=&
    (-ie)^{N}\int_{0}^{\infty} \!\ud T\, \prod_{i=1}^N\int^T_0d\tau_i \, \e^{-im^2 T}\!
    \int_{x(0) = x}^{x(T) = x'}\hspace{-1.5em}\mathcal{D}x(\tau) \notag\\ &\e^{i S_{\mathrm{B}}[a]-i\int^T_0\!d\tau\mathcal{J}\cdot x}\,\mathrm{Spin}[a](\tilde{F}_{1:N})\Big|_{\textrm{lin }N}\,,
\end{align}
where now $\mathrm{Spin}[a](\tilde{F}_{1:N})$ contains
\be\label{sf_start}
    \mathfrak{W}_{\eta}[a](\tilde{F}_{1:N}) 
    = 2^{-\frac{D}{2}}\oint_{\mathrm{A/P}}\mathcal{D}\psi(\tau)\, 
    \e^{i\widetilde{S}_{\mathrm{F}}[a]-i\sum_{j=1}^N \psi_{\eta}(\tau_j) \cdot \tilde{F}_j \cdot \psi_{\eta}(\tau_j)}\,.
\ee

\subsection{LSZ on the worldline}
\label{sec:lsz}
The connection from the propagator to corresponding amplitude can be facilitated through LSZ reduction. Although the exact asymptotic form of the truncated wave functions for the various kinds of backgrounds considered in Sec.~\ref{sec:back} may differ, the essential steps to perform LSZ on the worldline are similar; we use the approach of Refs.~\onlinecite{Bonocore:2020xuj, Mogull:2020sak}. 

For complex scalars LSZ for scattering can be performed by truncating both sides of the propagator in momentum space, extracting the residues, and then going on-shell for the scalar particle,
\be
\mathcal{A}_{N}^{p'p}= 
    \label{def:ampliude-N}
    \!\lim_{p'^2,p^2\rightarrow m^2-i0^\LCp}\!-(p'^2-m^2+i0^\LCp)(p^2-m^2+i0^\LCp)\mathcal{D}_{N}^{p'p}[a]\,,
\ee
where the Fourier transformed propagator is $\mathcal{D}_N^{p'p}[a] =\int\!\ud^4x'\ud^4 x\, \e^{ip'\cdot x'-ip \cdot x}\, \mathcal{D}^{x'x}_N[a]$. This holds for backgrounds that vanish in the asymptotic past and future. Fields whose background approach the constant valued $\lim_{x_0\to\infty}a(x)=a^\text{out}$ affect the following replacement: $\mathcal{D}_N^{p'p}[a]\to\mathcal{D}_N^{(p'+a^\text{out})p}[a]$.~\cite{Kibble:1975vz,Dinu:2012tj} Likewise for the in-state, in which $\lim_{x_0\to-\infty}a(x)=a^\text{in}$, one may use the replacement $\mathcal{D}_N^{p'p}[a]\to\mathcal{D}_N^{p'(p+a^\text{in})}[a]$. Note that in such cases the physical in and out momenta still lie on-shell, i.e. $p^{\prime 2},p^2=m^2$.

Next, let us illustrate how to truncate on the outgoing state first. Writing the propagator as $\mathcal{D}_N^{p'p}[a] =\int_{0}^{\infty}\!\ud T\, \e^{i({p'}^2-m^2 + i 0^\LCp)T} F(T)$, one may perform the following in the on-shell limit for $p'$:
\begin{align}\label{F_infty}
&-i({p'}^2-m^2 + i 0^\LCp)\mathcal{D}_N^{p'p}[a] \\
&=F(0) + \int_0^\infty\!\ud T \,  \e^{i({p'}^2-m^2 + i 0^\LCp)T}\frac{\ud}{\ud T} F(T)=F(\infty)\,.\notag
\end{align}
Truncation on the incoming state proceeds similarly; however, one must first extract a suitable propertime after some redefinitions. First, we can make the following change of variables~\cite{Mogull:2020sak} for the $N$ parameter integral variable
\be
\tau_0 \coloneqq \frac{1}{N}\sum_{i=1}^N\tau_i \;,
    \qquad
    \bar{\tau}_i\coloneqq \tau_i-\tau_0 \;,
\ee
with the collection of parameter integrals now expressible as
\be
    \prod_{i=1}^N\int_{0}^{\infty}\!\ud\tau_i
    =\int_{0}^{\infty}\!\ud\tau_0
    \prod_{i=1}^N\int_{-\infty}^{\infty}\!\ud \bar{\tau}_i \, \delta\Bigl(\sum_{j=1}^N\frac{\bar{\tau}_j}{N}\Bigr) \,.
\ee
What this does effectively is to furnish us with an additional Schwinger propertime-like parameter integral for the collective coordinate, $\tau_0$, that may be used as before with $T$ to now truncate the incoming state with $p$ going on-shell. Let us write the truncation on the incoming state as before with $F(\infty)\eqqcolon \int^\infty_0d\tau_0\,\e^{i(p^2-m^2+i0^\LCp)\tau_0} G(\tau_0)$ as
\be \label{G_infty}
    \mathcal{A}_{N}^{p'p}=-i(p^2-m^2+i0^\LCp)F(\infty)=G(\infty)\,,
\ee
which is determined in the same way as~\eqref{F_infty} with now $p$ going on-shell. And thus furnishing us with the desired scattering amplitude. The LSZ procedure can be visualized with Fig.~\ref{fig:lsz} below.
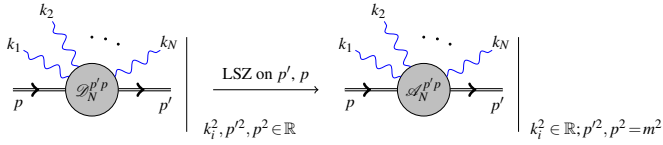
\begin{figure}
\scalebox{.7}{
     \begin{tikzpicture}
    \begin{scope}[decoration={
    markings,
    mark=at position 0.5 with {\arrow{>}}}
    ] 
            \draw[line width=0.2mm,double,postaction={decorate}] (-1.5,0) -- (-.5,0);
            \draw (-1.4,-.3) node {$p$};
            \draw[line width=0.2mm,double,postaction={decorate}] (.5,0) -- (1.5,0);
            \draw (1.4,-.3) node {$p'$};
            \draw[draw,fill=lightgray] (0,0) circle (0.5) node {$\mathcal{D}^{p'p}_{N}$};
            \draw[draw=blue, snake it] (-0.433013, 0.25) -- (-1.29904, 0.75);
            \draw (-1.47224, 0.85) node {$k_1$};
            \draw[draw=blue, snake it] (-0.25, 0.433013) -- (-0.75, 1.29904);
            \draw (-0.85, 1.5) node {$k_2$};
            \filldraw[black] (0.5, 0.866025) circle (.5pt);
            \filldraw[black] (0.258819, 0.965926) circle (.5pt);
            \filldraw[black] (0, 1) circle (.5pt);
            \draw[draw=blue, snake it] (0.433013, 0.25) -- (1.29904, 0.75);
            \draw (1.47224, 0.9) node {$k_N$};
            \draw (1.8,-.8) --  (1.8,1);
            \draw (3,-.75) node {$k_i^2,p'^2,p^2\!\in\!\mathbb{R}$};
            \hspace{-6em}
            \draw[line width=0.2mm, ->] (4.2,0) -- (6.2,0);
            \draw (5.2,.3) node {LSZ on $p'$, $p$};
            \draw[line width=0.2mm,double,postaction={decorate}] (-1.5+8.2,0) -- (-.5+8.2,0);
            \draw (-1.4+8.2,-.3) node {$p$};
            \draw[line width=0.2mm,double,postaction={decorate}] (.5+8.2,0) -- (1.5+8.2,0);
            \draw (1.4+8.2,-.3) node {$p'$};
            \draw[draw,fill=lightgray] (8.2,0) circle (0.5) node {$\mathcal{A}^{p'p}_{N}$};
            \draw[draw=blue, snake it] (-0.433013+8.2, 0.25) -- (-1.29904+8.2, 0.75);
            \draw (-1.47224+8.2, 0.85) node {$k_1$};
            \draw[draw=blue, snake it] (-0.25+8.2, 0.433013) -- (-0.75+8.2, 1.29904);
            \draw (-0.85+8.2, 1.5) node {$k_2$};
            \filldraw[black] (0.5+8.2, 0.866025) circle (.5pt);
            \filldraw[black] (0.258819+8.2, 0.965926) circle (.5pt);
            \filldraw[black] (8.2, 1) circle (.5pt);
            \draw[draw=blue, snake it] (0.433013+8.2, 0.25) -- (1.29904+8.2, 0.75);
            \draw (1.47224+8.2, 0.9) node {$k_N$};
            \draw (1.8+8.2,-.8) --  (1.8+8.2,1);
            \draw (3+8.8,-.75) node {\hspace{-2em}$k_i^2\in\mathbb{R};p'^2,p^2\!=\! m^2$};       
        \end{scope}
\end{tikzpicture}}
\caption{LSZ reduction process for complex scalars~\cite{Copinger:2024twl}. External matter lines represent fully dressed solutions in the background field; external photons (non-necessarily on shell) are labeled with $k_i$ momenta. LSZ sends the matter momenta on-shell, and yields an off-shell current/amplitude from the propagator for photons off/on-shell.}
\label{fig:lsz}
\end{figure} 

LSZ for the spinor case follows through in a very similar way. However, here we have a few extra terms to contend with--these turn out to be leading and subleading contributions to the amplitude, and they stem from the covariant derivative term acting on the spacetime kernel. To treat them judiciously one must start from the LSZ expression in coordinate space,
\begin{align}
\label{QED:amplitude:start}
   \mathcal{M}_{Ns's}^{p'p} =& i{\lim_{p'^2,p^2\rightarrow m^2-i0^\LCp}} \int \!\ud^{4}x'\ud^{4}x\,
    \e^{i(p'+a^\text{out})\cdot x'-ip\cdot x}
    \bar{u}_{s'}(p')\notag\\
    &(i\slashed{\partial}_{x'}-\slashed{a}^{\text{out}}-m)\mathcal{S}_{N}^{x'x}(-i\overleftarrow{\slashed{\partial}}_{x}-m)u_{s}(p)\,,
\end{align}
where we have shown for illustrative purposes the case in which $a^{\text{out}}$ does not vanish. One may likewise define the incoming state. We finally remark that after taking the derivatives and identifying the momentum space kernel, a similar form of LSZ reduction as was shown for the scalar case,~\eqref{def:ampliude-N}, in momentum space will emerge. Upon which a similar truncation can be accomplished on both ends in the worldline representation.

\section{\label{sec:planewaves}Plane waves}
\subsection{Complex Scalars}

Now that we have laid out all the tools necessary to compute all multiplicity $N-$photon scattering amplitudes in strong-field QED using the worldline formalism, let us begin evaluating backgrounds of interest towards ultra-intense laser modeling starting with the case of plane waves with background~\eqref{planewave}. Note that in this section we refer to the plane wave background field as $a(x)$, and we will do the same for the different background fields in coming sections, where we will also use $a(x)$. However, it is implied for each section that derived quantities such as propagators and amplitudes are assumed under the background field concerned for each section--here plane waves.

We begin with complex scalars, first evaluating the propagator in coordinate space given in~\eqref{D_N}. Let us first expand about the straight line: $x^{\mu}(\tau)= x^{\mu}+z^\mu\frac{\tau}{T}+q^{\mu}(\tau)$ with $z^\mu := x^{\prime\mu}-x^{\mu}$. This converts the inhomogeneous Dirichlet boundary conditions to homogeneous ones with $q(0)=q(T)$. In a plane wave background the worldline coordinate path integral in $q$ is Gaussian, albeit a hidden Gaussianity~\cite{AntonPlane,Edwards:2021vhg,Edwards:2021uif}. To extract this exactly-solvable behavior, we make use of an auxiliary variable technique~\cite{Schubert:2023gsl}. Essentially, the technique affects the following redefinition wherein a Lagrange multiplier $\chi(\tau)$ and auxiliary field $\xi(\tau)$ are introduced for the gauge field dependent term:
\begin{align}\label{hidden}
    &\e^{-i\int \ud\tau\, a(x^\LCp+z^\LCp\frac{\tau}{T}+q^\LCp) \cdot \dot{q}}
    =\\
    &\int\!\mathcal{D}\xi\mathcal{D}\chi\,
    \e^{i\int \ud\tau\, \big[\chi(\xi-q^\LCp)-a(x^\LCp+z^\LCp\frac{\tau}{T}+\xi)\cdot\dot{q}\big]}\notag\,.
\end{align}
The Lagrange multiplier and auxiliary field have the effect of moving the coordinate, $q$, dependence out of $a$, and onto the argument of the exponential. The key to this process is the fact the $\chi q^\LCp$ term will produce a term proportional to $n^\mu$ after integrating out $q$, which is orthogonal to both $a$ and also $n$, and hence supplying us with an exactly solvable Gaussian system. One need only employ the worldline Green function with homogeneous Dirichlet boundary conditions, i.e.,
$\langle q^{\mu}(\tau) q^{\nu}(\tau')\rangle = 2 i{\eta^{\mu\nu}} \Delta(\tau, \tau')$ with
\be\label{green0dbc}
\Delta_{ij} \coloneqq \Delta(\tau_{i},\tau_{j}) = \frac{1}{2}|\tau_{i}-\tau_{j}|-\frac{1}{2}(\tau_{i}+\tau_{j})+\frac{\tau_{i}\tau_{j}}{T}\,,
\ee
that satisfies $\partial_\tau^2\Delta_{\tau\tau'}=\delta(\tau-\tau')$ and $\Delta_{0\tau'}=\Delta_{T\tau'}=\Delta_{\tau 0}=\Delta_{\tau T}=0$. Completing the path integral in $q$ will now result in a term only linear in $\chi$ in the exponent, which will produce a delta function sending $\xi(\tau_i)\to 2i(\Delta_{ij}\varepsilon_j^\LCp+i\Delta_{ij}k_j^\LCp)$ and now all instances of the background field in the propagator can be written as
\be
a^\mu_i\equiv a^{\mu}(\tau_{i}) \equiv a^\mu\Bigl( x^{\LCp} + z^{\LCp}\frac{\tau_i}{T} +2\sum_{j=1}^{N}\big[-\Delta_{ij}k_{j}^{\LCp}+i\deld_{ij}\varepsilon_{j}^{\LCp}\big]  \Bigr)\,.
\ee
Dots here indicate propertime derivatives acting on either side, i.e., $\deld_{ij}=\partial_j\Delta_{ij}$ and $\ddel_{ij}=\partial_i\Delta_{ij}$. To achieve a simplified form from here in this section concerning plane waves we assume a gauge for the $N-$photons in which $\varepsilon^\LCp_i=0$. The final form of the coordinate space propagator becomes
\begin{align}
\mathcal{D}_{N}^{x'x}=&{i}
{(-e)^{N}}
\int_{0}^{\infty}\!
\frac{\ud T}{(4i\pi T)^{2}} \, \e^{-i\frac{z^{2}}{4T}}
{\prod_{i=1}^{N}
\int_{0}^{T} \ud\tau_{i}}\, 
\e^{-iM^{2}(a)T}
\Pn_{N}^{x'x}\notag\\
&\e^{-iz\cdot\llangle a\rrangle+ i\sum_{i=1}^{N}(x+\frac{z}{T}\tau_{i}-2I(\tau_i))\cdot k_{i}-i\sum_{i,j=1}^{N}\Delta_{ij}k_{i}\cdot k_{j}}\;,\\
{\bar{\mathfrak{P}}_{N}^{x'x}}=& 
i^{N}  \e^{\sum_{i=1}^{N}\varepsilon_{i}\cdot\frac{z}{T}+2\sum_{i=1}^{N}(\llangle a\rrangle-a_{i})\cdot\varepsilon_{i} }\notag\\
&\e^{i\sum_{i,j=1}^{N}[2i\ddel_{ij}\varepsilon_{i}\cdot k_{j} + \ddeld_{ij}\varepsilon_{i}\cdot\varepsilon_{j}]}{\Big|_{\mathrm{lin}\,N}\,.}
\end{align}
See Ref.~\cite{Copinger:2023ctz} for further details. In this form, one may see the appearance of a Kibble mass term~\cite{Kibble:1975vz} $M^{2}(a)\coloneqq m^{2}-\llangle a^{2}\rrangle+\llangle a\rrangle^{2}$, and the Bern Kosower exponent term in $\Pn_{N}^{x'x}$ that is independent of the background field. In addition, we have used the periodic integrals $\llangle f\rrangle\coloneqq T^{-1}\int_{0}^{T}\!\ud\tau \, f(\tau)$ and $I_{\mu}(\tau)\coloneqq\int_{0}^{\tau}\!\ud\tau' [ a_{\mu}(\tau')-\llangle a_{\mu}\rrangle]$.

To obtain the desired scattering amplitude one need only apply the steps as laid out in Sec.~\ref{sec:lsz}. The Fourier transform of the propagator does not present any major problems, and apart from one remaining integral (we use the initial light front time $x^\LCp$), all remaining integrals are exactly solvable. Because the plane wave gauge does not disappear at asymptotic infinity,~\eqref{planewave}, the end desired momentum space propagator is the shifted one, $\mathcal{D}^{\tilde{p}'p}_N$ for $\tilde{p}'=p'+a^\infty$. We find for the propagator in momentum space
\begin{align}\label{Dpp}
    &\mathcal{D}^{\tilde{p}'p}_{N} =  \int_N
    \e^{i({\tilde p}'^2-m^2) T+i({\tilde p}'+K-p)\cdot x+i(2{\tilde p}'+K)\cdot g}\\
    &\e^{-i\sum_{i,j=1}^{N}[\frac{1}{2}|\tau_i-\tau_j|k_i\cdot k_j-i\,\,\textrm{sgn}(\tau_i-\tau_j)\varepsilon_i\cdot k_j+\varepsilon_i\cdot\varepsilon_j\delta(\tau_i-\tau_j)]}\nonumber\\
    & \e^{-\int_{0}^{T}(2{\tilde p}' \cdot a(\tau)-a^2(\tau))\ud\tau-2\sum_{i=1}^{N}[\int_{0}^{\tau_i}k_{i}\cdot a(\tau)\ud\tau-i\varepsilon_i\cdot a(\tau_i)]}
    \bigg|_{\textrm{lin}\,N}\,,\nonumber
\end{align}
with implicit dependence now $a(x^\LCp + g^{\LCp} +(p'+p)^\LCp\tau-\sum_{i=1}^Nk_{i}^{\LCp}|\tau-\tau_i|)$. For brevity, we have written $g=\sum_{i=1}^{N}(k_i\tau_i-i\varepsilon_i)$ and $K=\sum_{i=1}^{N} k_i$. Also we have compactly written $\int_N = (-ie)^{N}\int \ud^4x\! \int_{0}^{\infty}\!\ud T \int_{0}^{T}\prod_{j=1}^{N} \ud\tau_j$.

Now we may perform LSZ truncation on both sides. We first truncate the outgoing state as~\eqref{F_infty}. However, to perform a truncation on the incoming state using~\eqref{G_infty}, one must first make the change of variables for the remaining $x^\LCp$ integral: $\bar{x}^\LCp := {x^\LCp+(p'+p+K)^\LCp\tau_0+ g^\LCp}$. This will further lead to the following implicit dependence for the gauge field:
$a(\bar\tau) \equiv a (
 \bar{x}^{\LCp}+(p'+p)^\LCp\,{\bar\tau}
    -\sum_{i=1}^{N}
    k_{i}^{\LCp}|{\bar\tau}-\bar{\tau}_i|)$.
Finally one may find for the scattering amplitude~\cite{Copinger:2023ctz}
 \begin{align}\label{Master:sQED:amplitude}
 	&\mathcal{A}^{p'p}_{N}=(-ie)^N\int\!\ud^4 x\,\e^{i(K+\tilde{p}'-p)\cdot x}
 \int_{-\infty}^{\infty}\prod_{i=1}^{N} \ud\tau_i\, 
 \delta\Bigl(\sum_{j=1}^N\frac{\tau_j}{N}\Bigr)
  \nonumber \\
 &\e^{-i\int_{-\infty}^{0}
 [2\tilde{p}' \cdot a(\tau)-a^{2}(\tau)]\ud\tau
 -i\int_{0}^{\infty}\!
 [2p'\cdot\delta a(\tau)-\delta a^{2}(\tau)]\ud\tau
 }\nonumber \\
 &\e^{i({\tilde p}' +p)\cdot g -2i\sum_{i=1}^{N}
 [\int_{-\infty}^{\tau_i}k_{i} \cdot a(\tau)\ud\tau-i\varepsilon_i\cdot a(\tau_i)]}     \nonumber \\
 &\e^{-i\sum_{i,j=1}^{N}\bigl(\frac{|\tau_i-\tau_j|}{2}k_i \cdot k_j-i\,\mathrm{sgn}(\tau_i-\tau_j)\varepsilon_i \cdot k_j+\delta(\tau_i-\tau_j)\varepsilon_i \cdot \varepsilon_j\bigr)}
 \bigg|_{\textrm{lin }N} \,,
 \end{align}
where $\delta a(x^{+}):=a(x^{+})-a^{\infty}$ and we have sent $\bar{\tau}\to\tau$. The above, in compact form, expresses any $N-$photon scattering for complex scalars. Let us now show the equivalent expression for spinors.

\subsection{Spinors}
To a large extent, the extension to spinors amounts to just the additional evaluation of the spin factor~\eqref{sf_start}. Therefore, let us first turn our attention towards its evaluation. The spin factor has an exactly solvable anti-periodic Green function for arbitrary background~\cite{Copinger:2023ctz} that may be used to evaluate the path integral; however, for the case of a plane wave background it takes on a simple form~\cite{Edwards:2021vhg}
\begin{align}
\mathfrak{G}(\tau,\tau')&=\mathrm{sgn}(\tau-\tau')
\Bigl[1-2\int_{\tau'}^{\tau}\!\ud\sigma \,f(\sigma)+2\Big(\int_{\tau'}^{\tau}\!\ud \sigma\, f(\sigma)\Big)^{2}\Bigr]
\notag\\&+T\llangle f\rrangle\Bigl[1-2\int_{\tau'}^{\tau}\!\ud \sigma\, f(\sigma)\Bigr]\,.
\end{align}
Such a solution is possible because for our background plane wave field, products of $f$ only to the quadratic order remain, even for different propertimes. The above satisfies
$(\frac{1}{2}\eta_{\mu\sigma}\partial_\tau+f_{\mu\sigma}{(\tau)})\mathfrak{G}^{\sigma\nu}{(\tau,\tau')}=\eta_{\mu}{}^{\nu}\,\delta(\tau-\tau')$, and where $\big\langle \psi^{\mu}(\tau) \psi^{\nu}(\tau')\big\rangle = \frac{1}{2}\mathfrak{G}^{\mu\nu}(\tau, \tau')$. The Green function satisfies anti-periodic boundary conditions, $\mathfrak{G}_{0\tau'}=-\mathfrak{G}_{T\tau'}$ and $\mathfrak{G}_{\tau 0}=-\mathfrak{G}_{\tau T}$. With the above solution, one may complete the square to find the following solution for the Grassmann path integral:
\begin{align}
\mathfrak{W}_{\eta}&(\tilde{F}_{1:N}) =
e^{-i\sum_{i=1}^N \frac{\delta}{\delta \theta (\tau_i)}\cdot \tilde{F}_{i} \cdot \frac{\delta}{\delta \theta (\tau_i)} }\e^{-\int_{0}^{T}\!\ud\tau \, [\eta \cdot f_\tau \cdot \eta +\j_\tau \cdot \eta]}
\notag\\&\e^{-\int_{0}^{T}\!\ud\tau
\ud\tau'[\eta \cdot f_\tau \cdot \mathfrak{G}_{\tau\tau'} \cdot \j_{\tau'}+\frac{1}{4}\j_\tau\cdot \mathfrak{G}_{\tau\tau'}\cdot \j_{\tau'}]}\Big|_{\j=0}\,.
\end{align}
The most important feature of this solution is that the `hidden Gaussianity' is still intact, for the spin factor only has $f$ field strength dependence whose implicit $q^\LCp$ dependence is shifted away by $\xi$ and the Lagrange multiplier $\chi$. Therefore all the previous analysis used for the complex scalar case holds here as well. Therefore let us go ahead and record the form of the spinor kernel in momentum space as
\begin{align}
    &\mathcal{K}^{p'p}_{N} =  \int_N
    \e^{i({\tilde p}'^2-m^2) T+i({\tilde p}'+K-p)\cdot x+i(2{\tilde p}'+K)\cdot g}\\
    &\e^{-i\sum_{i,j=1}^{N}[\frac{1}{2}|\tau_i-\tau_j|k_i\cdot k_j-i\,\,\textrm{sgn}(\tau_i-\tau_j)\varepsilon_i\cdot k_j+\varepsilon_i\cdot\varepsilon_j\delta(\tau_i-\tau_j)]}\nonumber\\
    & \e^{-\int_{0}^{T}(2{\tilde p}' \cdot a(\tau)-a^2(\tau))\ud\tau-2\sum_{i=1}^{N}[\int_{0}^{\tau_i}k_{i}\cdot a(\tau)\ud\tau-i\varepsilon_i\cdot a(\tau_i)]}
    \textrm{Spin}[a](\tilde{F}_{1:N})\bigg|_{\textrm{lin}\,N}\,,\nonumber
\end{align}
with similar implicit $N-$photon dependence in the arguments of all background fields. 

One may proceed as before to find the amplitude, however, we will now see that there are two subleading terms present, but they will vanish. First, let us write the amplitude for the case of our plane wave from~\eqref{QED:amplitude:start} as
\begin{align}
    \mathcal{M}_{Ns's}^{p'p} &=  
    \lim_{p'^2,p^2\rightarrow m^2}\notag
    \frac{i}{2m}\int d^{4}x'd^{4}x\, e^{i\tilde{p}'\cdot x'-ip\cdot x}\,\bar{u}_{s'}(p')
    \\&(p^{\prime 2}-m^2)\Bigl\{ \Bigl[-1
    +\frac{1}{2m} \delta \slashed{a}(x^{\prime +})\notag
    \Bigr]\,\mathcal{K}_N^{x'x}\\
    &+\frac{e}{2m}\sum_{i=1}^{N}\slashed{\varepsilon}_{i}e^{ik_{i}\cdot x'}\,\mathcal{K}_{N-1}^{x'x}\Bigr\}(p^2-m^2)u_{s}(p)\,.
\end{align}
The first subleading term is the one with $\slashed{a}(x^{\prime \LCp})$ factor. From a  Fourier transform we will pick up a $\delta(x^{\LCp}-x'^{\LCp}+2g^{\LCp}+2p^{\prime \LCp}T)$ factor, sending $\delta a(x^{\prime +}) \to \delta a(2Tp^{\prime +}+x^{\LCp}+2g^\LCp)$, which will vanish from outgoing truncation due to $T\to\infty$. The second subleading term is the one with the $\sum_{i=1}^{N}\slashed{\varepsilon}_{i}$ factor. Here the poles are not on the mass-shell $p^{\prime 2} -m^2$, but rather at $(p'+k_i)^2-m^2$, and will thus the factor will vanish after going on-shell. A similar finding was shown for the vacuum case~\cite{Ahmadiniaz:2021gsd}. With the remaining leading factor present in the amplitude, it is easy to see the ensuing momentum space propagator, and similar truncation as was demonstrated from the scalar case. Therefore we can go ahead and record the final expression for spinor scattering amplitude as~\cite{Copinger:2023ctz}
\begin{align}
 	&\mathcal{M}_{Ns's}^{p'p}=(-ie)^N\int\!\ud^4 x\,\e^{i(K+\tilde{p}'-p)\cdot x}
 \int_{-\infty}^{\infty}\prod_{i=1}^{N} \ud\tau_i\, 
 \delta\Bigl(\sum_{j=1}^N\frac{\tau_j}{N}\Bigr)
  \nonumber \\
 &\e^{-i\int_{-\infty}^{0}
 [2\tilde{p}' \cdot a(\tau)-a^{2}(\tau)]\ud\tau
 -i\int_{0}^{\infty}\!
 [2p'\cdot\delta a(\tau)-\delta a^{2}(\tau)]\ud\tau
 }\nonumber \\
 &\e^{i({\tilde p}' +p)\cdot g -2i\sum_{i=1}^{N}
 [\int_{-\infty}^{\tau_i}k_{i} \cdot a(\tau)\ud\tau-i\varepsilon_i\cdot a(\tau_i)]}     \nonumber \\
 &\e^{-i\sum_{i,j=1}^{N}\bigl(\frac{|\tau_i-\tau_j|}{2}k_i \cdot k_j-i\,\mathrm{sgn}(\tau_i-\tau_j)\varepsilon_i \cdot k_j+\delta(\tau_i-\tau_j)\varepsilon_i \cdot \varepsilon_j\bigr)}\nonumber\\
 &\frac{1}{2m}\bar{u}_{s'}(p')\mathrm{Spin}[a](\tilde{F}_{1:N})u_{s}(p)
    \Big|_{\textrm{lin}\,N}\,.
 \end{align}
Note the resemblance to the scalar case with addition of spin factor insertion.

\section{\label{sec:impulsive}Impulsive PP-Waves}

\subsection{Complex scalars}
Another important background field to consider for ultra-intense laser modeling is that of an impulsive plane-fronted waves with parallel propagation (PP)-wave, or ultra boosted arbitrary source distributions, whose gauge field we use is given in~\eqref{shockwave}. However, for this case in contrast to the plane wave case above where we used a Fourier transform, we make use of path integral boundary conditions that incorporate the Fourier transformation step to directly find the momentum space propagator. These boundary conditions are Robin boundary conditions and are mixed in the sense that
\be\label{robin}
    \frac{1}{2} \dot{x}_\mu(T)+ eA_\mu(x(T)) = p'_\mu \quad\text{and}\quad x(0)=x\,.
\ee
And the ensuing path integral in momentum space is
\be\label{robin_prop}
     \mathcal{D}^{p'p}_{N}[a] =\int_N \int_{\textrm{Robin}}\hspace{-1.5em}\mathcal{D}x(\tau)\, \e^{iS_{\mathcal{J}}[x(\tau);a]}\Big\rvert_{\textrm{lin}\,N}\,,
\ee
with modified worldlline action now reading
\be
S_{\mathcal{J}}[x(\tau);a] =S_\mathrm{B}[x(\tau);a] +p'\cdot x(T) - p\cdot x(0) -\int_{0}^{T}\!\ud\tau\,
   \mathcal{J}_\mu x^\mu(\tau)\,.
\ee
The novelty of this approach can be predicted as for the case of the impulsive PP-wave (and indeed the plane wave case too) there will a remaining integral over spacetime, as can be seen in the plane wave expression in~\eqref{Dpp} for $\int d^4x$ the initial coordinate, making it well suited for our background. To understand how to arrive at~\eqref{robin_prop} consider taking the Fourier transform of one side of the coordinate space propagator; this is will be the initial point of $x(0)=x$, and the $p\cdot x(0)$ term then just represents the Fourier transform. And for the other side one may vary the worldline action, $S_\textrm{B}$ over the endpoint $x(T)\rightarrow x(T)+\delta x_T$ and the subsquent integral over $\delta x_T$ will produce the other mixed boundary condition in~\eqref{robin}. See Ref.~\onlinecite{Copinger:2023ctz} for further details.

To evaluate~\eqref{robin_prop} in the impulsive PP-wave background one need only expand about the classical path of the light front coordinates, i.e., $x^{\LCpm}(\tau)=x^{\LCpm}_{\text{cl}}(\tau)+q^{\LCpm}(\tau)$ with $\ddot{x}^{\LCp}_{\text{cl}}=2\mathcal{J^{\LCp}}$, obeying the Robin boundary conditions. The hidden Gaussianity is still present here. Thus one may complete the two light front coordinate path integrals to find
\be
    \mathcal{D}^{p'p}_{N}[a] = \int_N \int_{\textrm{Robin}}\hspace{-1.5em}\mathcal{D}x^{\LCperp}
    \e^{iS_{\mathcal{J}}[(x^{\pm}_{\text{cl}},x^{\LCperp});0]+i\int_{0}^{T}\,\ud\tau\,
    {e(\dot{x}^{\LCp}_{\text{cl}})\delta(x^{\LCp}_{\text{cl}})\Phi(x^{\LCperp})
    }}
    \bigg\rvert_{\textrm{lin}\,N}\,.
\ee
With $\Phi(x^{\LCperp})$ representing an arbitrary function, the remaining path integral does not seem solvable; however the trick is to continue to a Fourier transformed quantity. First, we write the Wilson line factor as product over $\,\nroots$ roots of $x^\LCp_{\text{cl}}(\tau)=0$ as
\be\label{wilsonline}
\e^{i\int_{0}^{T}\ud\tau\,\dot{x}^{\LCp}_{\text{cl}}\delta(x^{\LCp}_{\text{cl}})\,e\,\Phi(x^{\LCperp}(\tau))}=\prod_{{j}=1}^{\nroots}
    \e^{ie\,\textrm{sgn}(\dot{x}^{\LCp}_{\text{cl}}(t_j))\,\Phi(x^{\LCperp}(t_j))}\,.
\ee
Then we may take the Fourier transform of the U$(1)$ factors~\cite{Tarasov:2019rfp,Adamo:2021jxz} to find
\be
    \e^{ie\, \textrm{sgn}(\dot{x}^\LCp)\Phi(x^{\LCperp})} =
    \int\! \hat{\ud}^2 r_\LCperp \,
    W(r_{\perp})\, 
    \e^{i\, \textrm{sgn}(\dot{x}^\LCp) r_{\perp} x^{\LCperp}}\,.
\ee
With this redefinition we can now see that for the remaining path integral we now have a Gaussian structure, albeit at the expense of introducing ordinary integrals over $W(r_{\perp})$.

\begin{figure}
\centering
\hspace{-3em}
\begin{subfigure}[b]{0.45\columnwidth}
\scalebox{.7}{
\begin{tikzpicture}
    \begin{scope}[decoration={
    markings,
    mark=at position 0.5 with {\arrow{>}}}
    ] 
            \draw[line width=0.4mm,red] (-2,2) -- (2,-2);
            \draw[line width=0.2mm,postaction={decorate}](-0.75,-2) -- (-.25,-1.5);
            \draw (-0.9,-2.2) node {$p$};
            \draw[line width=0.2mm,postaction={decorate}] (-.25,-1.5) -- (0,-.5);
            \draw[line width=0.2mm,postaction={decorate}](0,-.5) -- (.75,.5);
            \draw[line width=0.2mm,postaction={decorate}](.75,.5) -- (1,2);
            \draw (1.2,2.3) node {$p'$};
            \draw[->,draw=blue, snake it] (.25,-2)--(-.25,-1.5);
            \draw (.35,-2.2) node {$k_1$};
            \draw[->,draw=blue, snake it] (0,-.5) -- (-2,1.5);
            \draw (-2.2,1.3) node {$k_2$};
             \draw[draw=red,fill=orange] (.22,-0.22) circle (1.5pt);
             \draw[->,draw=blue, snake it] (2,-.75) -- (.75,.5);
             \draw (2,-1) node {$k_3$};
        \end{scope}
\end{tikzpicture}
}
  \end{subfigure}  
\begin{subfigure}[b]{0.45\columnwidth}
\scalebox{.7}{
\begin{tikzpicture}
    \begin{scope}[decoration={
    markings,
    mark=at position 0.5 with {\arrow{>}}}
    ] 
            \draw[line width=0.4mm,red] (-2,2) -- (2,-2);
            \draw[line width=0.2mm,postaction={decorate}](0,-2) -- (0,-1);
            \draw[line width=0.2mm,postaction={decorate}] (0,-1) -- (.5,0);
            \draw[line width=0.2mm,postaction={decorate}](.5,0) -- (-.75,.25);
            \draw[line width=0.2mm,postaction={decorate}](-.75,.25) -- (-1,2);
            \draw (-1,2.3) node {$p'$};
            \draw[->,draw=blue, snake it] (0,-1) --(-2,1);
            \draw (0,-2.2) node {$p$};
            \draw[->,draw=blue, snake it] (2,-1.5) -- (.5,0);
            \draw (-2.2,.8) node {$k_1$};
             \draw[draw=red,fill=orange] (.32,-0.32) circle (1.5pt);
             \draw[draw=red,fill=orange] (-.13,.13) circle (1.5pt);
             \draw[draw=red,fill=orange] (-.83,.83) circle (1.5pt);
             \draw[->,draw=blue, snake it] (-.75,.25) -- (-2,1.5);
             \draw (2,-1) node {$k_2$};
              \draw (-2.2,1.5) node {$k_3$};
        \end{scope}
\end{tikzpicture}
}
\end{subfigure}
\caption{Examples of the positivity constraint for $N=3$.~\cite{Copinger:2024twl} The red line represents the support of the PP-wave at $x_{\text{cl}}^\LCp$, and is on the $x^\LCpm$ plane. (Left): Example of $\,\nroots=1$ and meeting the positivity constraint. (Right): Example of a violation of the constraint with  $\,\nroots=3$ crossings of the shock.
}
\label{fig:positivity}
\end{figure}
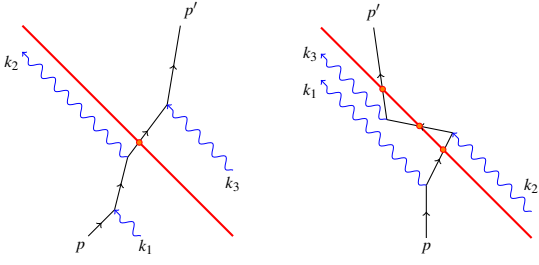

At this point one may perform the remaining path integral in $x^\LCperp$. However the range of solutions to $x_\text{cl}^\LCp=0$ is unconstrained and can become complicated. Let us therefore analyze a special case that admits a major simplification; this is one in which $\dot{x}_\text{cl}^\LCp$ is positive definite. The merit of this is that there is at most one solution of $\,\nroots$. Then the conditions under which $\dot{x}_\text{cl}^\LCp$ is positive definite are as follows: 1) $p^{\prime\LCp} >0$; this is actually already a condition for LSZ. And 2)
\be
     p^{\prime\LCp}+\sum_{i\in\mathcal{U}}k_i^\LCp >0  \quad \forall\;\; \mathcal{U}\subseteq \{1,2,3,...,N\}\,.
\ee
Both constraints we herein refer to as the `positivity constraint.' Constraint 2 occurs for example in multiple nonlinear Compton scattering. For a diagrammatic example of the positivity constraint see Fig.~\ref{fig:positivity}.

Let us now examine the propagator under the positivity constraint. In addition to the non-trivial solution with $\,\nroots=1$, there will be a trivial solution in which the shock is bypassed at $\,\nroots=0$. However the trivial solution will not survive LSZ and hence the physical amplitude we wish to show. Therefore let us simply define a new propagator that excludes what would be vanishing contributions after LSZ; this is 
\begin{align}\label{prpagator-vacuum-form0}
   {(-ie)}\widetilde{\mathcal{D}}^{p'p}_{N}=&
   {\int_{N+1}}
   \int {{\hat \ud}^3 k_{\scaleto{(N+1)\mathstrut}{5pt}}}
   W(k_{\scaleto{(N+1)\perp\mathstrut}{5pt}})\dot{x}^{\LCp}_{\text{cl}}(\tau_{\scaleto{(N+1)\mathstrut}{5pt}})\notag\\
   &\e^{iS_{\tilde{\mathcal{J}}}[\tilde{x}_{\text{cl}}(\tau);0]+ik_{\scaleto{(N+1)+\mathstrut}{5pt}}x_{\text{cl}}^{\LCp}(\tau_{\scaleto{(N+1)\mathstrut}{5pt}})}\Big\rvert_{\textrm{lin.}\,\varepsilon}\,,
\end{align}
where we have suggestively written $r_{\perp}=k_{\scaleto{(N+1)\perp\mathstrut}{5pt}}$. Notice now that the term with $W$ represents a scalar vertex with $\varepsilon_{{\scaleto{(N+1)\mathstrut}{5pt}} \mu }=n_{\mu}W(k_{\scaleto{(N+1)\perp\mathstrut}{5pt}})$. Now if we identify $k_{\mu}$ with $k_{\scaleto{(N+1)\mathstrut}{5pt}-}=0$, we can write the above as
\be\label{prpagator-vacuum-form}
    (-ie)\widetilde{\mathcal{D}}^{p'p}_{N}=
    \int\! {\hat \ud}^4 k_{\scaleto{(N+1)\mathstrut}{5pt}}
    {\hat \delta}(k_{\scaleto{(N+1)\mathstrut}{5pt}}^\LCp)\,
    D^{p'p}_{N+1}
    \,\Big\rvert_{\varepsilon_{\scaleto{(N+1)\mathstrut}{5pt}}
    =n_{\mu}W(k_{\scaleto{(N+1)\perp\mathstrut}{5pt}})}\,,
\ee
where $D^{p'p}_{N+1}$ is the free propagator without background field. Likewise one may using the LSZ scheme as shown before write the corresponding amplitude as~\cite{Copinger:2024twl}
\begin{align}
    (-ie)\mathcal{A}^{p'p}_{N}=
    \int\! {\hat \ud}^4 k_{\scaleto{(N+1)\mathstrut}{5pt}}
    {\hat \delta}(k_{\scaleto{(N+1)\mathstrut}{5pt}}^\LCp)\,
    A^{p'p}_{N+1}\Big\rvert_{\varepsilon_{\scaleto{(N+1)\mathstrut}{5pt}\mu}
    =
    n_{\mu}W(k_{\scaleto{(N+1)\perp\mathstrut}{5pt}})}\,,
\end{align}
with $A^{p'p}_{N+1}$ here being the free amplitude as well. What the above states is that in order to calculate the $N-$photon impulsive PP-wave amplitude under the positivity constraint, one need only compute the \textit{free} amplitude with $N+1-$photons. The Fourier integral over the U$(1)$ factor will remain though. This provides a major simplification for $N-$photon scattering at an arbitrary multiplicity on the impulsive PP-wave. Also, another identification in terms of a plane wave instead of the free case can be had; see Ref.~\onlinecite{Copinger:2024twl} for details.

\subsection{Spinors}

Let us now illustrate how a similar finding indeed arises for the case of spinors. We begin with kernel in momentum space with Robin boundary conditions, here written as
\be
\mathcal{K}^{p'p}_{N}
    = \int_N \int_{\textrm{Robin}}\hspace{-1.5em}\mathcal{D}x(\tau)\, \e^{iS_{\mathcal{J}}[x(\tau);a]}
    \textrm{Spin}[a](\tilde{F}_{1:N})
    \Big|_{\textrm{lin}\,N}\,.
\ee
We now specialize to the case of an impulsive PP-wave background. And for much of the following discussions it suffices to examine the action of the spin factor. Thus, the action here takes the form
\be
    \widetilde{S}_{\mathrm{F}}=
    \int_{0}^{T}\!\ud\tau\Big[\frac{i}{2}\psi\cdot\dot{\psi}+2ie\delta(x^{\LCp})\psi_\eta^{\LCp}\psi_\eta^\LCperp\partial_\LCperp\Phi(x^{\LCperp})\Big]\,.
\ee
Now when we expand about the classical path, where as before with the complex scalar case, the above action becomes
\be
\widetilde{S}_{\mathrm{F}}
    =\frac{i}{2}\int_{0}^{T}\!\ud \tau\,\psi\cdot\dot{\psi}+\sum_{j}^{\nroots}
    \frac{2ie}{|\dot{x}_{\text{cl}}^{\LCp}(t_{j})|}\psi_\eta^\LCp(t_{j})
    \psi_\eta^\LCperp(t_{j})
     {\partial_\LCperp \Phi(x^{\LCperp}(t_{j}))}\,.
\ee
The PP-wave only provides support at a given $t_j$. Next we use the fact that for Grassmann variables $\psi_\eta^\LCp(t_{j})^2=0$ to find for the exponential of the action
\be
    \e^{i\widetilde{S}_{\mathrm{F}}}
    =\e^{\frac{i}{2}\int_{0}^{T}\!\ud\tau\psi\cdot\dot{\psi}}\prod_{j}^{\nroots}
    \Bigl[1-\frac{2e}{|\dot{x}_{\text{cl}}^{\LCp}(t_{j})|}\psi_\eta^\LCp(t_{j})
    \psi_\eta^\LCperp(t_{j})
    \partial_\LCperp\Phi(x^{\LCperp}(t_{j}))\Bigr]\,.
\ee
We desire at this point to perform the same kind of Fourier transform so as to complete the path integral. To do so let us couple the above to the Wilson line, c.f.~\eqref{wilsonline}, to find
\begin{align}
    &\e^{i\widetilde{S}_{\mathrm{F}}}\prod_{j}^{\nroots}
    \e^{ie\,\mathrm{sgn}(\dot{x}_{\text{cl}}^{\LCp}(t_{j}))\Phi(x^{\LCperp}(t_{j}))}\\
    &=\e^{\frac{i}{2}\int_{0}^{T}\!\ud\tau\psi\cdot\dot{\psi}}
    \prod_{j}^{\nroots}
    \Bigl[
    1+\frac{2i}{\dot{x}_{\text{cl}}^{\LCp}(t_{j})}\psi_\eta^\LCp(t_{j})
    \psi_\eta^\LCperp(t_{j})
    {\partial_\LCperp}\Bigr]
    \e^{ie\,\mathrm{sgn}(\dot{x}_{\text{cl}}^{\LCp}(t_{j}))\Phi(x^{\LCperp}(t_{j}))}.
    \notag
\end{align}
Now we can use the same Fourier transform as was used for just the exponential factor. Examining just one $j$ we can now write
\be
 {\int \hat{\ud}^{2}r_{j\perp} \Big[1-\frac{2}{|\dot{x}_{\text{cl}}^{\LCp}(t_{j})|}\psi_\eta^\LCp(t_{j})
    \psi_\eta^\LCperp(t_{j})r_{j\LCperp}\Big] W(r_{j\perp}) \e^{i\, \mathrm{sgn}(x^{+}) r_{j\perp} x^{\perp}}}\,,
\ee
which indeed combines the scalar and spinor parts into the U$(1)$ factor, and have as before an action that is Gaussian in $x^\LCperp$.

Analogous arguments hold for the spinor case as well when we expand over $\,\nroots$ under the positivity constraint, in that only the $\,\nroots=1$ case will remain after LSZ. Thus we can go ahead and immediately write down the kernel that will undergo LSZ as~\cite{Copinger:2024twl}
\begin{align}
    &(-ie)\widetilde{\mathcal{K}}^{p'p}_{N} =\int_{N+1}
    \int{\hat{\ud}}^3 k_{\scaleto{(N+1)\mathstrut}{5pt}}
    \e^{iS_{\tilde{\mathcal{J}}}[\tilde{x}_{\text{cl}}(\tau);0]+ik_{\scaleto{(N+1)\LCp\mathstrut}{5pt}} x^\LCp_{\text{cl}}(\tau_{\scaleto{(N+1)\mathstrut}{5pt}})}\notag\\
    &\bigl[W(k_{\scaleto{(N+1)\perp\mathstrut}{5pt}})\dot{x}^\LCp_{\text{cl}}(\tau_{\scaleto{(N+1)\mathstrut}{5pt}})\bigr]
    \mathrm{symb}^{-1}\mathfrak{W}_{\eta}[a](\tilde{F}_{1:N}) 
    \Big{|}_{\textrm{lin}\,N}\,,
\end{align}
where we have went ahead and denoted the modified kernel with a tilde, which represents the remaining terms after LSZ. Let us look at the term above in brackets coupled with the background field part of the spin factor; this is 
\be
    W(k_{\scaleto{(N+1)\perp\mathstrut}{5pt}})\dot{x}^\LCp_{\text{cl}}(\tau_{\scaleto{(N+1)\mathstrut}{5pt}})-2W(k_{\scaleto{(N+1)\perp\mathstrut}{5pt}})\psi_\eta^\LCp(\tau_{\scaleto{(N+1)\mathstrut}{5pt}})
    \psi_\eta(\tau_{\scaleto{(N+1)\mathstrut}{5pt}})\cdot k_{\scaleto{(N+1)\mathstrut}{5pt}}\,.
\ee
This resembles a worldline photon vertex operator for scalars and spinors. Therefore let us label $r_{\perp}=k_{\scaleto{(N+1)\perp\mathstrut}{5pt}}$ and $\varepsilon_{{\scaleto{(N+1)\mathstrut}{5pt}} \mu}=n_{\mu}W(k_{\scaleto{(N+1)\perp\mathstrut}{5pt}})$ as before. This then permits the following distinction, c.f.~\eqref{prpagator-vacuum-form}:
\be
    (-ie)\widetilde{\mathcal{K}}^{p'p}_{N}=\int\!    {\hat \ud}^4 k_{\scaleto{(N+1)\mathstrut}{5pt}} \, 
    {\hat{\delta}}(k_{\scaleto{(N+1)\mathstrut}{5pt}}^\LCp)K^{p'p}_{N+1}\Big\rvert_{\varepsilon_{\mu\scaleto{(N+1)\mathstrut}{5pt}}\rightarrow n_{\mu}W(k_{\scaleto{(N+1)\perp\mathstrut}{5pt}})}\,.
\ee
$K$ of course represents the free $N-$photon dressed propagator. In order to make the connection to the amplitude, however, we must contend with the subleading factors in the LSZ expression. Fortunately, these fall away by a very similar reason as was argued in the plane wave case. Namely here, for the subleading factor proportional to the background field since there is no support at asymptotic infinity for $\delta(x^\LCp)$, it will vanish after truncating the outgoing state. Also, for the subleading factor proportional to the sum over $N-$photons, here too as well the pole will lie away from the mass shell, and it too will vanish once going on-shell. Therefore, we can go ahead and make the analogous arguement for the spinor case as well in that~\cite{Copinger:2024twl}
\be
    (-ie)\mathcal{M}^{p'p}_{Ns's}= 
    \int\!    {\hat \ud}^4 k_{\scaleto{(N+1)\mathstrut}{5pt}} \, 
    \hat{\delta}
    (k_{\scaleto{(N+1)\mathstrut}{5pt}}^\LCp)M^{p'p}_{(N+1)s's}\Big\rvert_{\varepsilon_{\mu\scaleto{(N+1)\mathstrut}{5pt}}\rightarrow n_{\mu}W(k_{\scaleto{(N+1)\perp\mathstrut}{5pt}})}\,,
\ee
where naturally $M^{p'p}_{(N+1)s's}$ describes the scattering amplitude for spinor without a background field.

\section{\label{sec:nonnull}Non-null fields}

Having examined scattering amplitudes for both the cases of plane waves and impulsive PP-waves, we now turn our attention to the case of non-null fields~\cite{Heinzl:2016kzb,Mackenroth:2018rtp,Mackenroth:2020pct}, with background gauge field given by~\eqref{nonnull}. As many of the discussions to follow in this section follow for spinors as well as scalars, the spinor discussions add unnecessary technical details for the physics we wish to highlight, therefore in this section we will turn our attention exclusively to the case of complex scalars. We provide only a high-level overview here--see, however, Ref.~\onlinecite{non_null}, for in-depth discussions on both cases. 

We will find that much of the discussion here parallels the case of the plane wave background. However, unlike the previous cases non-null backgrounds do not admit exact solutions; thus we employ a perturbative scheme about the non-null parameter, $\rho^2$.\footnote{$\rho^2$ by itself is not a physical dimensionless parameter. But we will show the relevant parameter scales as $\rho^2/p^{\prime{\tri}}$ after calculating the amplitude.} However, under the expansion we can find a solution as a function of functional derivatives. To show this let us first write down the auxiliary and Lagrange multiplier function in the non-null gauge as
\be
     \e^{-i\int \ud\tau\, a(x^{\tri}(\tau)) \cdot \dot{q}}  = \int \mathcal{D}\xi\mathcal{D}\chi\,
\e^{i\int \ud\tau\, \big[\chi(\xi-q^{\tri})-a(\widehat{x}+\xi(\tau))\cdot\dot{q}\big]}\,.
\ee
Unlike the plane wave case,~\eqref{hidden}, we no longer benefit from a hidden Gaussianity since $\rho^2\neq0$. We can proceed as before with the plane wave case in Sec.~\ref{sec:planewaves} by expanding about the straight line. While isolated instances of $x^\LCp$ for the plane wave case are replaced by $x^{\tri}$ for the non-null case, we further pick up a distinctly new term at the exponential level after integrating out the fluctuations, $q$:
\be
    -i\int\!\ud \tau \ud\tau' \, \rho^2 \chi(\tau)\chi(\tau') \Delta(\tau, \tau')\,.
\ee
With the plane wave case we had but a linear term that gave us a delta function. We use that solution to serve as an expansion about small $\rho^2$. To show this let us gather the $\chi$ dependent terms and express the above term in terms of functional derivatives,
\begin{align}\label{int_chi}
    &\int \mathcal{D} \chi(\tau) \, \e^{-i \rho^{2}\int\!\ud \tau \ud\tau' \, \chi(\tau)\chi(\tau') \Delta(\tau, \tau') + i \int \ud\tau \chi(\tau) \mathcal{I}(\tau)}\notag\\
    &=\e^{i \rho^{2}\int \ud \tau' \ud \tau'' \Delta(\tau', \tau'') \frac{\delta}{\delta \xi(\tau')} \frac{\delta}{\delta \xi(\tau'')}}\delta[\mathcal{I}(\tau)]  \,,
\end{align}
where $\mathcal{I}(\tau) =  \xi(\tau)  - 2i\sum_{j=1}^{N} [\deld(\tau, \tau_{j}) \varepsilon_{j}^{\tri} + i\Delta(\tau, \tau_{j})k_{j}^{\tri}]$. In this way we can see that as $\rho^2$ goes to zero the plane wave case is reproduced. We can actually use this definition to evaluate the path integral in its entirety, albeit in functional derivative form, to find for the $N-$photon coordinate space propagator
\begin{align}
    & \mathcal{D}_{N}^{x'x} = i(-ie)^{N}\int_{0}^{\infty} dT \, (4\pi iT)^{-2}  \e^{i \rho^{2}\int \ud \tau' \ud \tau'' \Delta(\tau', \tau'') \frac{\delta}{\delta \xi(\tau')} \frac{\delta}{\delta \xi(\tau'')}} \notag\\
    &\Big\lbrace \e^{ -iM^{2}(a)T-i \frac{z^{2}}{4T} - i z \cdot \llangle a \rrangle + i\sum_{j=1}^{N} \big[ k_{j} \cdot (x + z \frac{\tau_{j}}{T}) - i \varepsilon_{j} \cdot \frac{z}{T} \big]} \notag \\
 & \e^{ 2\sum_{j=1}^{N} \big[ \big(\llangle a \rrangle - a(\tau_{j})\big)\cdot \varepsilon_{j} - i I(\tau_{j})\cdot k_{j} \big]
-i\sum_{i,j=1}^{N}[\Delta_{ij}k_{i}\cdot k_{j} - 2i\ddel_{ij}\varepsilon_{i}\cdot k_{j} - \ddeld_{ij} \varepsilon_{i}\cdot\varepsilon_{j}]
} \notag\\
&\Big\rbrace \Big|_{\textrm{lin}\,N}^{\xi(\tau) \rightarrow 2i\sum_{j=1}^{N} [\deld(\tau, \tau_{j}) \varepsilon_{j}^{\tri} + i\Delta(\tau, \tau_{j})k_{j}^{\tri}]}\,.
\end{align}
After taking the functional derivatives it can be seen the implicit dependence given in the background field will be
\be
    a\big(x^{\tri} + z^{\tri}\frac{\tau}{T} + 2i\sum_{j=1}^{N} [\deld(\tau, \tau_{j}) \varepsilon_{j}^{\tri} + i\Delta(\tau, \tau_{j})k_{j}^{\tri}]\big)\,.
\ee
The above equations are valid for any $\rho^2$. However, at this point let us go ahead and analyze the smallest non-vanishing contributions coming from the functional derivative term to $\mathcal{O}(\rho^2)$. Even so, we remark that there still is an implicit $x^{\tri}$ dependence in the background field. Thus this is not a true perturbative treatment, and is a \textit{partial resummation} as $\rho^2$ is kept to all orders in the background field~\cite{Heinzl:2016kzb}.

To arrive at the amplitude using the approach in Sec.~\ref{sec:lsz}, we will first need the momentum space propagator. The Fourier integrals can be exactly evaluated, and we eventually find that
\begin{align}
    &\mathcal{D}^{\tilde{p}'p}_{N} =  \int_N (1+\rho^2 \Delta_1)\notag
    \e^{i({\tilde p}'^2-m^2) T+i({\tilde p}'+K-p)\cdot x+i(2{\tilde p}'+K)\cdot g}\\
    &\e^{-i\sum_{i,j=1}^{N}[\frac{1}{2}|\tau_i-\tau_j|k_i\cdot k_j-i\,\,\textrm{sgn}(\tau_i-\tau_j)\varepsilon_i\cdot k_j+\varepsilon_i\cdot\varepsilon_j\delta(\tau_i-\tau_j)]}\nonumber\\
    & \e^{-\int_{0}^{T}(2{\tilde p}' \cdot a(\tau)-a^2(\tau))\ud\tau-2\sum_{i=1}^{N}[\int_{0}^{\tau_i}k_{i}\cdot a(\tau)\ud\tau-i\varepsilon_i\cdot a(\tau_i)]}
    \bigg|_{\textrm{lin}\,N}\,,
\end{align}
where the first order correction to the plane wave case reads
\be\label{delta1}
    \Delta_1=-2i\int_{0}^{T}d\tau d\tau'|\tau-\tau'|P(\tau)\cdot a'(\tau)P(\tau')\cdot a'(\tau')\,,
\ee
and where $P(\tau)=\tilde{p}'-a(\tau)+\sum_{i=1}^{N}[\theta(\tau_{i}-\tau)k_{i}-\delta(\tau_{i}-\tau)i\varepsilon_{i}]$. Dashes over the background field indicate derivatives with respect to the coordinate, $x$ (here the classical background solution), i.e. $a'(\tau)=\frac{\partial}{\partial x_{\text{cl}}^\LCp(\tau)}a(\tau)$.

Fortunately the same steps as used in Sec.~\ref{sec:lsz} to arrive to the amplitude for the plane wave case can be applied here as well. We find that~\cite{non_null}
\begin{align}\label{nonnulA}
 	&\mathcal{A}^{p'p}_{N}=(-ie)^N\int\!\ud^4 x\,\e^{i(K+\tilde{p}'-p)\cdot x}
 \int_{-\infty}^{\infty}\prod_{i=1}^{N} \ud\tau_i\, 
 \delta\Bigl(\sum_{j=1}^N\frac{\tau_j}{N}\Bigr)
  \nonumber \\
 &\e^{-i\int_{-\infty}^{0}
 [2\tilde{p}' \cdot a(\tau)-a^{2}(\tau)]\ud\tau
 -i\int_{0}^{\infty}\!
 [2p'\cdot\delta a(\tau)-\delta a^{2}(\tau)]\ud\tau
 }\nonumber \\
 &\e^{i({\tilde p}' +p)\cdot g -2i\sum_{i=1}^{N}
 [\int_{-\infty}^{\tau_i}k_{i} \cdot a(\tau)\ud\tau-i\varepsilon_i\cdot a(\tau_i)]} 
 (1+\rho^2\Delta_1)\nonumber \\
 &\e^{-i\sum_{i,j=1}^{N}\bigl(\frac{|\tau_i-\tau_j|}{2}k_i \cdot k_j-i\,\mathrm{sgn}(\tau_i-\tau_j)\varepsilon_i \cdot k_j+\delta(\tau_i-\tau_j)\varepsilon_i \cdot \varepsilon_j\bigr)}
 \bigg|_{\textrm{lin }N} \,,
 \end{align}
with implicit $N-$photon dependence in the non-null field being now
$a(x^{\triangle}+(p^{\prime\triangle}+p^{\triangle})\tau-\sum_{i=1}^{N}k_{i}^{\triangle}|\tau-\tau_{i}|)$. And we take the integration ranges within $\Delta_1$ in~\eqref{delta1} to now be over $\mathbb{R}$ as $\int^T_0\to\int^\infty_{-\infty}$.

A diagrammaic approach to the case of a non-null background in a partial resummed scheme, as we have too demonstrated here, leads to an expansion over the dimensionless parameter $\rho^2/p^{\prime{\tri}}$ or $\rho^2/p^{{\tri}}$ for scattering amplitudes~\cite{Heinzl:2016kzb}. The same dimensionless expansion parameter is here as well. In the $\rho^2\Delta_1$, examining $\Delta_1$, one can see that a redefinition from propertime coordinates to spacetime $x^{\tri}$, one will acquire for every $\rho^2$ term, an accompanying $1/p^{\prime{\tri}}$ factor. This is easy to see for the simplest case of the wavefunction--with $N=0$ and truncation on just one side. There the implicit dependence in the background field will go to $x_{cl}^{\triangle}(\tau)=x^{\triangle}+(p^{\prime\triangle}+p^{\triangle})\tau$, where we will acquire a $1/(p^{\prime\triangle}+p^{\triangle})$ factor after redefinition. Higher order in $N$ will introduce cuts in the propertime and augment the integration region over the $k_i$; however similar redefinitions are possible upon applying conservation of momentum. In practice, however, it is more straightforward to redefine the properime integrals in the expression of the amplitude,~\eqref{nonnulA}, to show the known expansion scheme (or to simply use the dimensional $\rho^2$ with understanding that the true expansion parameter is $\rho^2/p^{\prime{\tri}}$), than it is to convert the worldline form to space time form.

\section{\label{sec:low}Homogeneous Fields with Low-energy photons}

The last case we will examine benefits from a tremendous simplification allowing us to probe higher multiplicity with ease; it relies on a low-energy approximation, Sec.~\ref{sec:back}(c), for the $N-$photons~\cite{ChrisLow}. Again here as in the previous section we restrict our attention to the case of complex scalars; the arguments to follow may be applied to the case of spinors without difficulty, and are done so in Ref.~\onlinecite{low_energy}. It is simplest to first show the low-energy approximation on the loop or rather from an $N-$photon multilinear expansion of the one-loop effective action. This object for complex scalars is 
\be
    \Gamma_N=(-ie)^N\int^\infty_0\!\frac{dT}{T}\,e^{-im^2T}\oint\!\mathcal{D}x(\tau)e^{iS_{\mathrm{B}}[a]}\prod_{i=1}^{N} V[\varepsilon_{i}, k_i]\,,
\ee
and has periodic boundary conditions for the path integral. The vertex \textit{under a standard gauge choice} is the same given in~\eqref{A_gamma} but under periodic boundary conditions. Then, a single vertex under the low-energy approximation keeping terms to only $\mathcal{O}(k_i^\mu)$ results in
\be
    V[\tilde{F}_i]=\!\int^T_0\!d\tau \,\frac{1}{2}x(\tau) \cdot \tilde{F}_i \cdot \dot{x}(\tau)+\mathcal{O}((k_i^\mu)^2)\,,
\ee
where we remind that $\tilde{F}^{\mu\nu}_i\coloneqq i[k^{\mu}_i\varepsilon^{\nu}_i-k^{\nu}_i\varepsilon^{\mu}_i]$ and $\tilde{f}_i=e\tilde{F}_i$. The vertex now resembles a constant field, providing a major simplification. While this identification is known on the loop~\cite{Edwards:2018vjd,Ahmadiniaz:2023jwd,MishaLow}, how one might extend it to the line? For if we use the same standard gauge given in~\eqref{A_gamma} this will lead to boundary contributions already at zeroth order. We therefore define our external photons here, in contrast to previous sections, under a Fock-Schwinger gauge, which in the low-energy limit becomes $\tilde{a}_\mu=-\frac{1}{2}\tilde{f}_{\mu\nu}(x^\nu-\hat{x}^\nu)$, or a constant field~\cite{low_energy}, where $\hat{x}^\nu$ is a reference point. Then, even on the line one may write
\be
    V^{x'x}[\tilde{F}_i]=\!\int^T_0\!d\tau \,\frac{1}{2}(x(\tau)-\hat{x}) \cdot \tilde{F}_i \cdot \dot{x}(\tau)+\mathcal{O}((k_i^\mu)^2)\,,
\ee
which for any reference point is \textit{linear} in $\tilde{F}$. With the above identification we may actually go ahead and gather the $N-$photon dependence into a compact form by writing
\be
    \prod_{i=1}^{N} (-ie)^NV[\tilde{F}_i]=\exp\Bigl[-i\!\int_{0}^{T}\!d\tau\,\tilde{a} \cdot \dot{x}\Bigr]\Big|_{\text{lin}\,N}
\ee
with effective homogeneous field $\tilde{a}^\mu = \sum_{i=1}^N \tilde{a}_i^\mu=-\frac{1}{2}\sum_{i=1}^N\tilde{f}^{\mu\nu}(x-\hat{x})_\nu$. With the above identification
we may go ahead and write down the full low-energy $N-$photon effective action and propagator as~\cite{low_energy}
\be
    \mathcal{D}_{N}^{x'x}=\mathcal{D}^{x'x}[a+\tilde{a}]\Big|_{\text{lin}\,N}\,,\;
    \Gamma_{N}=\Gamma[a+\tilde{a}]\Big|_{\text{lin}\,N}\,.
\ee
What these say is that all one needs to do to evaluate the full low-energy $N-$photon effective action or propagator is evaluate the one with background field plus a homogeneous background field composed of the superposition over $N$-photon fields strengths, and then apply the linearization operator. This is a major simplification over the ordinary expressions. However, one must still evaluate the linearization operator, and this process provides the last computational difficulty and thus is the remaining subject of this section. 

One more simplification can be had if we select as our background field a homogeneous field, for then the combined field too will be homogeneous. In the context of laser field modeling one may select a CCF, e.g. $f_{\mu\nu}=n_\mu a_\nu-n_\nu a_\nu$; however, we elect to keep the homogeneous background field arbitrary as it does not hinder our analysis. Even so, let us briefly outline how to convert to a CCF. Like the plane wave case since $n^\mu$ is orthogonal to both itself and $a^\mu$, products of three or more $f$ vanish. The final form form for the effective action and propagator is still applicable, but the Green functions and other background field dependent object will take on a simplified form.

\subsection{Effective action}
Let us first do the linear expansion for the effective action. Since the combined homogeneous field and low-energy $N-$photon field are still just another homogeneous field we already know the solution to the effective action as the Euler-Heisenberg effective action~\cite{heisenberg2006consequences}, which we write in a suggestive way:
\be
    \Gamma[a+\tilde{a}]=-i\mathcal{V}\!\int^\infty_0\!\frac{dT}{(4\pi)^2T^3}\,\e^{-im^2T}\,K[f+\tilde{f}]\,,
\ee
where the \textit{Schwinger propertime} kernel reads
\begin{align}\label{K_determinant}
    K[f+\tilde{f}]&= \frac{\mathrm{Det}{}^{-\frac{1}{2}} [ \hat{\partial}_{\tau}^{2} 
    -2(f+ \tilde{f})\hat{\partial}_{\tau}]  }{\mathrm{Det}^{-\frac{1}{2}} [ \hat{\partial}_{\tau}^{2}]}\\
    &=\det{}^{-\frac{1}{2}} \Big[ \frac{\sinh(Z+\tilde{Z})}{Z+\tilde{Z}}\Big]\,.
\end{align}
with $Z\coloneqq fT$ and $\tilde{Z} \coloneqq \tilde{f} T$. $\mathcal{V}$ is the spacetime volume factor. Writing the Schwinger propertime kernel using the $\ln \det=\tr \ln$ trick,
\be
    K[f+ \tilde{f}]=K[f] \e^{ -\frac{1}{2} \mathrm{Tr} \ln\big[ 1 - 2(\hat{\partial}_{\tau}^{2} 
	-2f\hat{\partial}_{\tau})^{-1}  \tilde{f} \hat{\partial}_{\tau}  \big] }\,,
\ee
serves as a suitable platform to expand around small $\tilde{f}$ in the log. Notice the inverse operator contained above. This is a worldline Green function; however, here for period boundary conditions. It satisfies $(\partial_\tau^2 -2f\partial_\tau)\mathcal{G}_{\tau\tau'}=2\delta(\tau-\tau')-2/T$; the $2/T$ factor accounts for the zero mode in the spectrum~\cite{Schmidt:1993rk}. The solution to the Green function~\cite{Schmidt:1993rk} in Minkowski spacetime is
\be\label{greenpbc}
    \mathcal{G}_{ij}=
    \frac{T}{2Z^{2}}\Bigl(1-\frac{Z}{\sinh(Z)}e^{-Z\dot{G}_{ij}}-Z\dot{G}_{ij}\Bigr)\,,
\ee
with free Green function given by $G_{ij}=|\tau_i-\tau_j|-T^{-1}(\tau_i-\tau_j)^2$. It satisfies the boundary conditions $\mathcal{G}_{0\tau'}=\mathcal{G}_{T\tau'}$ and $\mathcal{G}_{\tau 0}=\mathcal{G}_{\tau T}$. With the Green function we can then write logarithmic expansion as follows:
\be\label{K_expanded}
    K[f+ \tilde{f}]= K[f]\e^{\sum_{n=1}^\infty \frac{(-i)^n}{2n}\mathcal{I}^{(n)}}\,,
\ee
where
\be
     \mathcal{I}^{(n)}=\!i^n\prod_{i=1}^n 
    \int^T_0\!d\tau_i \,\mathrm{tr} (\dot{\mathcal{G}}_{12}\tilde{f} \dot{\mathcal{G}}_{23}\tilde{f} \!... \,\dot{\mathcal{G}}_{n1}\tilde{f} )\,.
\ee
Since $K_N[f]=K[f+ \tilde{f}] |_{\text{lin}\,N}$, to calculate $K_N[f]$ it is clear that one only needs to truncate the sum in the above expression to $N$ and then to apply the linearization operator take all permutations involving $N$ $\tilde{f}$. Since $\tilde{f}=\sum_{i=1}^N \tilde{f}_i$ applying the linearization operator on $N$ $\tilde{f}$ will generate all the symmetric combinations of $\tilde{f}_i$. Let us show the two lowest orders found from the expression above. They are
\begin{align}
    K_1[f]=&\frac{e}{2} K[f]\int^T_0\!d\tau_1\mathrm{tr} [\dot{\mathcal{G}}_{11}\tilde{f}_1]\,,\\
    K_2[f]=&\frac{e^2}{2}K[f]\int^T_0 \!d\tau_1d\tau_2\,\Bigl[\mathrm{tr} (\dot{\mathcal{G}}_{11}\tilde{f}_{1})\mathrm{tr} (\dot{\mathcal{G}}_{22}\tilde{f}_{2})\notag\\
    &+\mathrm{tr} (\dot{\mathcal{G}}_{12}\tilde{f}_{1}\dot{\mathcal{G}}_{21}\tilde{f}_{2})  \Bigr]\,.
\end{align}
Further, one must still compete the parameter integrals of $\tau_i$; these can lead to lengthy expressions, therefore, we keep the final form in integral form as above. However, see Ref.~\onlinecite{low_energy} for instructions on how to compute the various integrals introduced in this section.

\subsection{Propagator}

Let us next address the scalar propagator. Its solution is well-known in homogeneous fields~\cite{PhysRev.82.664}:
\be
    D^{x'x}[a + \tilde{a}]=-i\!\int^\infty_0\!\frac{dT}{(4\pi T)^2}e^{-im^2T}K^{x'x}[f + \tilde{f}]\,,
\ee
where the Schwinger propertime kernel here is
\begin{align}\label{Kxx}
	K^{x'x}[f+ \tilde{f}]&=K[f+ \tilde{f}]\e^{-i\frac{z^{2}}{4T}+iz\cdot\frac{(Z+\tilde{Z})^2}{T^{4}}\oDel(f+ \tilde{f}){}^{\circ}\cdot z} \notag\\
	&=K[F+ f]\e^{-\frac{i}{4T}z\cdot(Z+ \tilde{Z})\coth(Z+ \tilde{Z})\cdot z}\,.
\end{align} 
We have taken the liberty of expressing the propertime kernel in terms of the worldline Green function, $\Del{}_{\tau\tau'}$. Also, here we use $\oDel_\tau\coloneqq\int^T_0d\tau'\Del{}_{\tau'\tau}$. The Green function here obeys homogeneous Dirichlet boundary condition, and is the extension of the free Green function given in~\eqref{green0dbc} to incorporate homogeneous fields. It has the following solution in terms of the periodic Green function,~\eqref{greenpbc}~\cite{Ahmadiniaz:2020wlm}
\be
    2 \Del{}_{ij}(f) \eqqcolon 2 \Del{}_{ij} = \G_{ij} - \G_{i0} - \G_{0j} + \G_{00}\,,
\ee
where we use an abbreviation for the background field case. The Green function here obeys the same boundary conditions as the free case. 

Also we see the same determinant factor as was present in the effective action. They are in fact the same, that is the determinant whose functional form is given in~\eqref{K_determinant}, for periodic boundary conditions for the effective action and for homogeneous Dirichlet boundary conditions here. We can see this in the expanded form for the determinant under periodic boundary conditions,~\eqref{K_expanded}, where notice that for $\mathcal{I}^{(n)}$ one may replace each $\dot{\G}_{ij}\to\dot{\G}_{ij}-\dot{\G}_{i0}$ and hence $\dDel_{ij}$.

Now all that is needed is to find a suitable expansion as was done previously for the effective action. We may use the previous expansion for $K[f+\tilde{f}]$ given in~\eqref{K_expanded}. However we will need now also to expand the coordinate dependent part of the action, and we do so by expanding $\Del{}_{ij}(f+\tilde{f})$ about $\tilde{f}$. This is formally a geometric series in $\tilde{f}$ and reads
\begin{align}
    &\Del(f+ \tilde{f})_{\tau\tau'}=\Del{}_{\tau\tau'}+\sum_{n=1}^\infty(-i)^n\Del{}^{(n)}_{\tau\tau'}\,,\notag\\
	&\Del{}^{(n)}_{\tau\tau'}=(2i)^n\prod_{i=1}^n 
    \int^T_0\!d\tau_i \,\Del{}_{\tau 1}\tilde{f} \dDel_{12}\tilde{f} \!... \,\dDel_{n\tau'}\,.
\end{align} 
This then leads to the compact final expression~\cite{low_energy}
\begin{align}
	&K^{x'x}[f+ \tilde{f}]=K^{x'x}[f] \\
	&\e^{iz\cdot \frac{\tilde{Z}Z\!+\!Z\tilde{Z}\!+\!\tilde{Z}^2}{T^4}\oDelo\cdot z+\sum_{n=1}^\infty (-i)^n\bigl[\frac{1}{2n} \mathcal{I}^{(n)}+iz\cdot \frac{(Z+ \tilde{Z})^2}{T^4}\oDel{}^{(n)}{}^{\circ}\cdot z \bigr]}\notag\,.
\end{align}
Let us go ahead and show the first order correction as
\begin{align}
    K^{x'x}_1[f]&=K^{x'x}[f]\,\mathcal{I}^{(1)x'x}\Big|_{\text{lin}1}\,,\\
    \mathcal{I}^{(1)x'x}&=\frac{i}{T^4}z\cdot 
	 (\tilde{Z}Z+Z\tilde{Z})\oDelo -iZ^2\oDel{}^{(1)}{}^{\circ}\cdot z-\frac{i}{2}\mathcal{I}^{(1)} \,.
\end{align}
And the second order correction is
\begin{align}
    K^{x'x}_2[f]=&K^{x'x}\,\bigl[\tfrac{1}{2!}\bigl(\mathcal{I}^{(1)x'x}\bigr)^2+\mathcal{I}^{(2)x'x}\bigr]\Big|_{\text{lin}2}\,,\\
    \mathcal{I}^{(2)x'x}=& \frac{i}{T^4}z\cdot\tilde{Z}^2\oDelo-i (\tilde{Z}Z+Z\tilde{Z})\oDel{}^{(1)}{}^{\circ} 
    - Z^2\oDel{}^{(2)}{}^{\circ}\cdot z\notag\\
    &-\frac{1}{4} \mathcal{I}^{(2)} \,.
\end{align}

Unlike in the previous sections, here we keep the propagator in its coordinate form. One may perform a Fourier transform of the propagator after evaluation~\eqref{Kxx}, or before completing the path integral resulting in an augmented Green function; both forms are shown in Ref.~\onlinecite{low_energy}. However, the extension to an amplitude by LSZ, and hence the formulae in Sec.~\ref{sec:lsz}, is ill-defined in a homogeneous field. Instead to arrive at the corresponding scattering amplitude, one may join the ends of the propagator with in and out going wavefunctions in their asymptotic limit. These steps are beyond the scope of this work, and will be presented in full in Ref.~\onlinecite{low_energy}.

\section{\label{sec:conclusions}Conclusions}

We have explored several different kinds of background field configurations with relevance towards ultra-intense lasers, which include plane waves, impulsive PP-waves, non-null backgrounds, and homogeneous fields with low-energy photons. With the aid of the worldline formalism dressed with $N-$photon vertices we were able to find compact expressions for Master Formulae for propagators and scattering amplitudes. This would be hard--or even impossible--using the standard Feynman diagrammatic approach to SFQED. For the cases of the plane waves and impulsive PP-waves we benefited from a hidden Gaussianity that enabled us to write exact expressions for any multiplicity of $N-$photons. Further, for the impulsive PP-wave case we were able to show how the amplitude in the background field for $N-$photons was related to another amplitude without background field (or the free case) for $(N+1)-$photons. Master formula for a non-null background for complex scalars was found for small $\rho^2$, enabling a close comparison to the case of a plane wave, and potential phenomenological value. Finally we discussed the case of a homogeneous background coupled to low-energy photons. The low-energy photon vertex operator resembled that of a homogeneous field as well, which provided a major simplification.

There are several important extensions to the work presented here.
Since compact formulae are presented by virtue of the worldline formalism, one immediate and important next step would be providing concrete expressions measurable at current laser facilities for finite $N-$photon scattering amplitudes. To this end, cross sections of the given amplitudes would be specified. Furthermore, for the case of low-energy photons coupled to a homogeneous (CCF) field on the line, rigorously defining an LSZ procedure would be necessary to furnish the associated amplitudes. It would be interesting for future experiments to determine how high-multiplicity tree-level scattering scales with the non-linearity parameters, and whether a breakdown of our perturbative understanding is present in such instances. Also, it is of interest to see if the ideas presented here have extension to classical gravity or even non-Abelian fields. For the latter this would be of particular interest for the case of low-energy photons. Finally, there is significant interest in extending our understanding not just in higher multiplicity, but rather also in higher loops and their resummation.

\begin{acknowledgments}
PC and JPE are supported by the EPSRC Standard Grant  EP/X02413X/1, and KR is supported by the EPSRC Standard Grant EP/X024199/1. IA would like to acknowledge support from the University Research Studentship (URS) at University of Plymouth.
\end{acknowledgments}

\bibliography{references}

\end{document}